\documentclass[
 aps,
 amsmath,amssymb,
 reprint,
]{revtex4-1}

\usepackage{graphicx}
\usepackage{color}
\usepackage{bm}
\usepackage{dcolumn}
\usepackage[english]{babel}
\usepackage[utf8]{inputenc}
\usepackage{textcomp}
\usepackage[T1]{fontenc}
\usepackage[colorlinks=true,linkcolor=blue,citecolor=blue,urlcolor=blue, anchorcolor=red, filecolor=red, menucolor=red]{hyperref}
\usepackage{float}

\begin{document}

\title{Synchrotron radiation from ultrahigh-intensity laser-plasma interactions and competition with Bremsstrahlung in thin foil targets}

\author{B. Martinez}
\email{bertrand.martinez@tecnico.ulisboa.pt}
\altaffiliation{Present address: GoLP/Instituto de Plasmas e Fus{\~a}o Nuclear, Instituto Superior T{\'e}cnico, Universidade de Lisboa,
Lisbon, Portugal}
\affiliation{CEA, DAM, DIF, F-91297 Arpajon, France}
\affiliation{CELIA, UMR 5107,Universit\'e de Bordeaux-CNRS-CEA, 33405 Talence, France}

\author{E. d'Humi\`eres}
\affiliation{CELIA, UMR 5107,Universit\'e de Bordeaux-CNRS-CEA, 33405 Talence, France}

\author{L. Gremillet}
\email{laurent.gremillet@cea.fr}
\affiliation{CEA, DAM, DIF, F-91297 Arpajon, France}
\affiliation{Universit\'e Paris-Saclay, CEA, LMCE, 91680 Bruy\`eres-le-Ch\^atel, France}

\begin{abstract}
By means of particle-in-cell numerical simulations, we investigate the emission of high-energy photons in laser-plasma interactions under ultrahigh-intensity conditions relevant to multi-petawatt laser systems. We first examine the characteristics of synchrotron radiation from laser-driven plasmas of varying density and size. In particular, we show and explain the dependence of the angular distribution of the radiated photons on the transparency or opacity of the plasma. We then study the competition of the synchrotron and Bremsstrahlung emissions in copper foil targets irradiated by $10^{22}\,\rm W\,cm^{-2}$, 50~fs laser pulses. Synchrotron emission is observed to be maximized for target thicknesses of a few 10~nm, close to the relativistic transparency threshold, and to be superseded by Bremsstrahlung in targets a few $\mu$m thick. At their best efficiency, both mechanisms are found to radiate about one percent of the laser energy into photons with energies above $10\,\rm keV$. Their energy and angular spectra are thoroughly analyzed in light of the ultrafast target dynamics.  
\end{abstract}

\maketitle

\section{Introduction}

The interaction of a high-intensity ($I_L \ge 10^{18}\,\rm W cm^{-2}$) laser pulse with an initially solid material sample leads to a significant fraction of the laser energy (from a few $\%$ to $\sim 50\,\%$) being converted into relativistic electrons~\cite{PRLWilks1992, PRELefebvre1997, PPCFDavies2009, NFKemp2014}. While propagating through the dense inner region of the illuminated target (or through a secondary convertor target), these electrons can radiate part of their kinetic energy, either directly through Bremsstrahlung \cite{PRLKmetec1992, APLGahn1998, RSIPerry1999, PoPHatchett2000, APLEdwards2002, NJPGaly2007, PoPCompant2012}, or indirectly through radiative relaxation of excited atomic states \cite{RSIYasuike2001, PREStephens2004, PRLSefkow2011}. The former process gives rise to continuous broadband photon spectra that extend up to the maximum fast electron energy, whereas the latter yields discrete spectra determined by atomic line transitions. Both types of fast-electron-induced radiation can serve for high-resolution flash radiography of dense objects \cite{PRLGlinec2005, PHPRavasio2008, PoPPark2008, PREBranbrink2009, PoPWestover2010, PoPCourtois2011, PoPJarrott2014, PREAntonelli17}, absorption spectroscopy of heated plasmas \cite{PRLAudebert2005, HEDPLecherbourg2007}, or characterization of the fast-electron distribution \cite{PREPisani2000, PRLSantala2000, PREMartinolli2006, PoPChen2009, IEEEMeadowcroft2012, NJPZulick2013}. In addition, laser-driven high-energy Bremsstrahlung photon sources have been exploited to trigger photonuclear reactions \cite{PRLCowan2000, ScienceLedingham2003, EPLSchwoerer2003, PRLPomerantz2014, PoPWang2017}, as well as to generate unprecedented dense electron-positron pair beams through the Bethe-Heitler process in high-$Z$ thick targets \cite{PRLLiang1998, APLGahn2000, PRLChen2009, NCSarri2015, PoPWilliams2016}. 

At the extreme laser intensities ($I_L \gtrsim 10^{22}\,\rm W cm^{-2}$) achievable at forthcoming multi-petawatt laser systems~\cite{OLSung2017, OLZeng2017, MREWeber2017, HPLSEPapadopoulos2016, CLEOPapadopoulos2019}, copious emission of energetic photons can also originate from direct laser-electron interaction, that is, through nonlinear inverse Compton scattering of the laser light by relativistic electrons
\cite{PRLZhidkov2002, PREKoga2004}. In the strong-field limit (such that $a_L \equiv eE_L/m_e c \omega_L \gg 1$, with $E_L$ as the laser field strength, $\omega_L$ the laser frequency, $c$ the light speed, $m_e$ the electron mass and $e$ the elementary charge) where the quasi-stationary field approximation holds, this mechanism is analogous to synchrotron emission~\cite{RMPErber1966, PPCFKirk2009, RMPDiPiazza2012}, and its efficiency is determined by the electron quantum parameter
\begin{equation}
\chi_e = \frac{\gamma}{E_S} \left[ \left(\mathbf{E}_\perp + \mathbf{v}\times \mathbf{B} \right)^2 + E_\parallel^2/\gamma^2 \right]^{1/2} \,,
\end{equation}
where $\mathbf{v}$ and $\gamma$ are the electron velocity and Lorentz factor, $(\mathbf{E},\mathbf{B})$ is the electromagnetic field, and $E_S = m_e^2c^3/\hbar e = 1.3\times 10^{18}\,\rm Vm^{-1}$ the Schwinger field. The subscripts $_\parallel$ and $_\perp$ denote vector components parallel or normal to $\mathbf{v}$, respectively. When $\chi_e$ approaches unity (specifically when $\chi_e \gtrsim 0.1$), the average photon energy is a significant fraction of the electron kinetic energy and the emission should be treated quantum mechanically~\cite{RMPErber1966, PPCFKirk2009, RMPDiPiazza2012}.

Nonlinear inverse Compton/synchrotron emission can be mediated not only by the laser field but also by the strong quasistatic fields possibly induced during the laser-plasma interaction~\cite{PRLStark2016}, or even by the self-fields of colliding, high-density electron-positron pair beams~\cite{PRABDelGaudio2019}. All-optical generation of $\gamma$-ray photons (with energies in the $\sim 0.1-10\,\rm MeV$ range) through nonlinear inverse Compton scattering was first achieved by making collide a relativistic ($>100\,\rm MeV$) electron beam issued from a plasma-wakefield accelerator with a moderately relativistic ($I_L \sim 10^{19}\,\rm W cm^{-2}$) femtosecond laser pulse~\cite{NPPhuoc2012, PRLSarri2014}. In those pioneering experiments, however, the quantum parameter was too low ($\chi_e \lesssim 0.01$) for the electron dynamics to be sizably affected by the radiation. Only recently, through the use of more intense lasers ($I_L \sim 4\times 10^{20}\,\rm W cm^{-2}$) and higher-energy ($\sim 2\,\rm GeV$) wakefield-driven electron beams, have the first measurements of inverse Compton scattering in the radiation reaction regime ($\chi_e \gtrsim 0.2$) been carried out, providing evidence for substantial (up to $\sim 30\,\%$) radiation-induced electron energy losses~\cite{PRXCole2018, PRXPoder2018}.

The above scenario of laser-electron-beam collisions has attracted most of the experimental interest so far, because it allows the quantum parameter to be maximized at fixed laser intensity~\cite{PRLBlackburn2014, PRLVranic2014, JPPRidgers2017}, and thus offers a promising testbed for quantum radiation reaction models~\cite{PoPSokolov2009, PREBulanov2011, PPCFMackenroth2013, PRADiPiazza18} under well-controlled conditions. Yet this setup usually involves two synchronized powerful laser pulses (one for generating the electron beam and one for colliding with it), and so poses strong experimental constraints.  
Therefore, in view of future experiments at ELI-class facilities, it remains worthwhile to further investigate the properties of laser-driven radiation in a simpler configuration whereby a single ultraintense laser pulse interacts with a plasma layer. According to previous works, significant ($\gtrsim 1\,\%$) energy conversion efficiency into high-energy radiation may be achieved at laser intensities $\gtrsim 10^{22}\,\rm W cm^{-2}$ in near-critical-density plasmas~\cite{PoPBrady2014, PoPNerush2014, PoPWang2015a, PoPChang2017}.

In this context, it is important to determine the interaction conditions leading to synchrotron emission prevailing over Bremsstrahlung, and therefore the scaling of the two competing radiation processes with the target parameters. This problem has as yet only been touched upon, although there is an increasing number of particle-in-cell (PIC) codes
that can self-consistently describe both synchrotron radiation and Bremsstrahlung~\cite{PoPPandit2012, CLFWard2014, EPJDWan2017, PPCFVyskocil2018, HPLSEWu2018, PoPMartinez2019}. Notably, Pandit~\emph{et al.}~\cite{PoPPandit2012} found that synchrotron emission dominates in $5\,\rm \mu m$ thick Cu targets irradiated at intensities exceeding $\sim 10^{22}\,\rm W cm^{-2}$. More recently, Wan~\emph{et al.}~\cite{EPJDWan2017} showed dominance of synchrotron emission at $I_L \ge 10^{21}\,\rm W cm^{-2}$ (resp. $\ge 10^{22}\,\rm W cm^{-2}$) in $1\,\rm \mu m$ thick Al (resp. Au) targets. Still, these studies did not examine the influence of the target thickness on the radiation, with the notable exception of~Vysko{\v c}il \emph{et al.}~\cite{PPCFVyskocil2018} who looked into the variations in the Bremsstrahlung spectrum from solid foils made of various materials and driven at $I_L \simeq 3\times10^{21}-10^{23}\,\rm W cm^{-2}$; their investigation, however, was restricted to micrometric thicknesses and, while apparently included in their simulations, synchrotron emission was not commented upon. These previous works motivate us to further scrutinize the competition between synchrotron and Bremsstrahlung radiation in targets driven by femtosecond laser pulses, which will be the main and final objective of this study. Before that, we will reexamine the dependence of laser-driven synchrotron radiation on the plasma parameters. We will restrict ourselves to the case of a not-so-extreme ($I_L=10^{22}\,\rm W cm^{-2}$) laser intensity, relevant to ELI-class facilities during their first years of operation.

This article is structured as follows. In Sec.~\ref{sec:sync_unif_plasmas}, we present a series of somewhat idealized PIC simulations, using planar laser waves, in order to characterize the synchrotron emission from plasmas of varying density and thickness. These simulations are designed to give insight into the processes at play in more realistic simulations of the laser-induced radiation from copper foil targets, as reported in Sec.~\ref{sec:comp_brem_sync}. There, the Bremsstrahlung and synchrotron emissions are analyzed as a function of the target thickness, and shown to strongly depend on the transparent or opaque character of the plasma. Specifically, synchrotron emission attains its maximum for target thicknesses of a few 10~nm, close to the relativistic transparency threshold, and becomes dominated by Bremsstrahlung in targets a few $\mu$m thick. Our concluding remarks are gathered in Sec.~\ref{sec:conclusions}.

\section{Synchrotron emission in uniform plasmas} \label{sec:sync_unif_plasmas}

In this section, by means of two-dimensional (2-D) PIC simulations, we characterize the laser-driven synchrotron radiation from uniform plasmas of varying density and thickness, giving rise to either in-depth penetration of the laser wave or to its absorption/reflection at the plasma boundary. Our main purpose is to identify distinct, density-dependent regimes of synchrotron emission, in light of which the experimentally relevant, integrated simulations of Sec.~\ref{sec:comp_brem_sync} will be analyzed.

\subsection{Numerical setup and modeling}

Our simulations have been performed using the \textsc{calder} PIC code~\cite{NFLefebvre2003, JPCSLobet2016, PPCFMartinez2018}.
The laser pulse is modeled as an electromagnetic plane wave of wavelength $\lambda_L \equiv 2\pi c/\omega_L = 1\,\rm \mu m$, peak intensity $I_L=10^{22}\,\rm W cm^{-2}$ ($a_L = 85$), linearly polarized along the $y$ axis, and propagating in the $+x$ direction. Unless otherwise stated, it has a constant temporal profile, preceded by a two-cycle-long ($6.6\,\rm fs$) linear ramp. The irradiated plasma slab is made of fully ionized carbon ions (C$^{6+}$) and electrons of uniform density profile. Introducing the critical density $n_c \equiv m_e \epsilon_0 \omega_L^2/e^2 \simeq 1.1\times 10^{21}\,\rm cm^{-3}$ ($\epsilon_0$ is the vacuum permittivity), the initial electron density is set to either $n_{e0}=17n_c$ or $100n_c$, leading, respectively, to relativistic self-induced transparency (RSIT) or opacity of the plasma. The density profile is either of finite length ($l=1\,\rm \mu m$) or `semi-infinite', i.e., long enough to prevent both the laser pulse and the accelerated particles from reaching its rear boundary over the time span of the simulations ($t \simeq 150\,\rm fs$). The time origin ($t=0$) is chosen to be when the laser peak intensity hits the (sharp) plasma front boundary, located at $x=16\,\rm \mu m$.

The 2D domain comprises $4800 \times 400$ cells, with cell size $\Delta x = \Delta y = \lambda_L/60$. Each cell initially contains 10 macro-particles per plasma species. The time step is $\Delta t = 0.6\Delta x$. The boundary conditions for both fields and particles are taken to be absorbing in the $x$ direction and periodic in the $y$ direction. Coulomb collisions between charged particles~\cite{PoPPerez2012} and synchrotron radiation~\cite{JPCSLobet2016} are described. The synchrotron module implemented in \textsc{calder} combines a continuous radiation reaction model~\cite{PoPSokolov2009} for electrons with $\chi_e \leq 10^{-3}$ and a Monte Carlo quantum model~\cite{PPCFDuclous2011} for electrons with $\chi_e \ge 10^{-3}$. The chosen threshold value between the two regimes is quite arbitrary, yet ensures that the quantum regime is accurately captured. Bremsstrahlung is not modeled in this Section. Since we do not describe the subsequent interaction of the radiated photons with the plasma particles or the electromagnetic fields, they are not advanced on the simulation grid but their properties are stored for post-processing. 

\subsection{Relativistically undercritical plasma}
\label{subsec:undercritical}

\begin{figure}[t]
\centering
\includegraphics[width=0.49\textwidth]{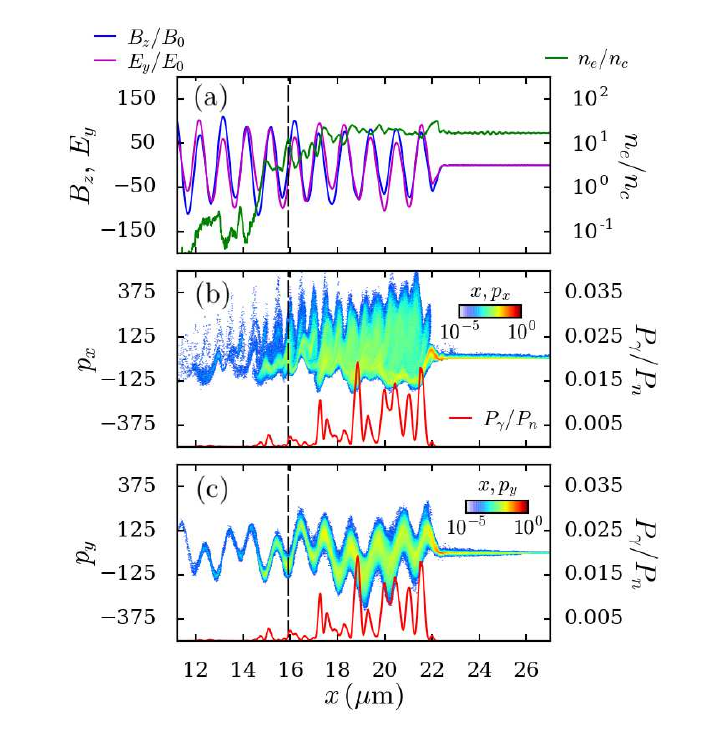}
\caption{Interaction of a semi-infinite, electromagnetic plane wave ($I_L=10^{22}\,\rm W cm^{-2}$) with a semi-infinite, relativistically undercritical C$^{6+}$ plasma ($n_{e0}=17n_c$). (a) Longitudinal lineouts of the $E_y$ (purple) and $B_z$ (blue) field components, and of the electron density $n_e$ (green). (b) $x-p_x$ electron phase space (averaged along $y$). (c) $x-p_y$ electron phase space (averaged over $y$). In (b) and (c) the red curve plots a longitudinal lineout of the synchrotron radiated power density ($P_\gamma$). All quantities are recorded $36 ,\rm fs$ after the on-target laser peak.}
\label{fig:uc_plasmas_interaction}
\end{figure}

We first consider the case of a semi-infinite plasma of density $n_{e0}=17n_c$. The main features of the laser-plasma interaction and ensuing high-energy radiation are illustrated at time $t=36\,\rm fs$ in Figs.~\ref{fig:uc_plasmas_interaction}(a-c). Figure~\ref{fig:uc_plasmas_interaction}(a) plots lineouts of the $E_y$ and $B_z$ field components (in units of $E_0 = m_e c \omega_L/e = 3.2\times 10^{12}\,\rm V m^{-1}$ and $B_0=m_e\omega_L/e = 1.1\times 10^4\,\rm T$, respectively) as well as of the electron density (in units of $n_c$). One can see that the laser wave has then travelled a few $\rm \mu m$ through the plasma (the vertical dashed curve indicates the vacuum/plasma interface). Albeit modulated by the laser ponderomotive force and the induced plasma waves, the electron density profile keeps an average value close to its initial value, as expected in the RSIT regime~\cite{PoPWeng2012}.

Figures~\ref{fig:uc_plasmas_interaction}(b,c) show the $x-p_x$ and $x-p_y$ electron phase spaces (both averaged along $y$). In each panel is overlaid the longitudinal profile of the radiated power density $P_\gamma$, normalized to $P_n = (2/3) \alpha_f n_c m_e c^2/\tau_C \simeq 3.4\times 10^{26}\,\rm W cm^{-3}$ ($\alpha_f$ is the fine structure constant and $\tau_C=\hbar/m_e c^2$ the Compton time). The emission is seen to take place  throughout the irradiated plasma region, in which the electrons have been accelerated to ultrarelativistic (longitudinal and transverse) momenta. The $x-p_x$ phase space exhibits the usual forward-moving, high-energy (up to $p_x/m_ec \approx 500$) electron jets spatially modulated at $\lambda_L/2$, but also a denser electron return current accelerated at $\vert p_x\vert /m_ec \approx 100-200$. Those counterstreaming electrons are first pushed forward in the rising edge of the laser wave before getting pulled back by the charge separation field, as analyzed by Debayle \emph{et al.}~\cite{NJPDebayle2017}. The laser front moves at a velocity $v_f/c \approx 0.47$, somewhat lower than that predicted ($v_f/c \simeq 0.56$) from Ref.~\onlinecite{PoPWeng2012}, probably as a result of mobile ions that favor electron compression (up to $n_e \approx 40n_c$) at the laser head. Transverse electron momenta as  high as $\vert p_y \vert/m_e c \approx 300$ are observed in Fig.~\ref{fig:uc_plasmas_interaction}(c), which may seem surprising since one expects $\vert p_y/ \vert m_e c \le \vert E_y\vert /E_0$ for an electromagnetic plane wave propagating in a dissipation-free plasma. In the present case, however, synchrotron radiation causes dissipation and, more importantly, the laser profile is subject to transverse modulations (not shown), leading to local field maxima $\vert B_z \vert/B_0 \approx 120$, so that the transverse canonical momentum is no longer conserved.

\begin{figure}[!t]
\centering
\includegraphics[width=0.48\textwidth]{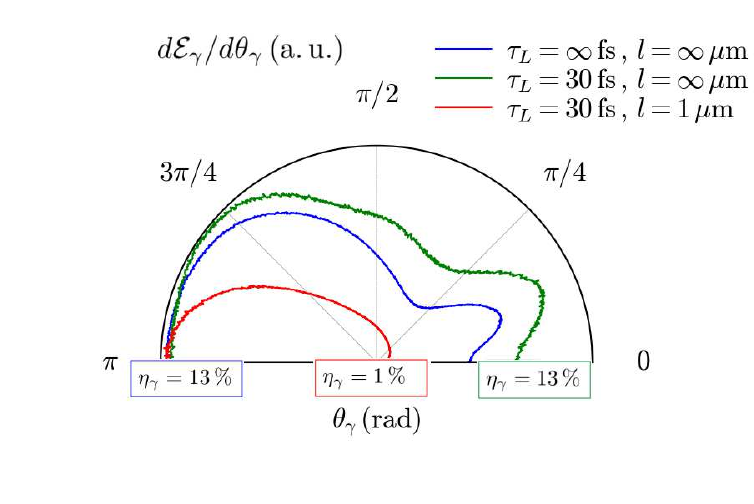}
\caption{Angle-resolved synchrotron radiated energy ($d\mathcal{E}_\gamma/d\theta_\gamma$) in a uniform C$^{6+}$ plasma slab with $n_e = 17n_c$, irradiated at a $10^{22}\,\rm Wcm^{-2}$ laser intensity. Three cases are considered: a semi-infinite laser wave in a semi-infinite plasma (blue), a  $30\,\rm fs$  laser pulse in a semi-infinite plasma (green), and a $30\,\rm fs$ laser pulse in a $1\,\rm \mu m$ plasma (red). Angles are defined as $\theta_\gamma =\arccos\left(k_{\gamma,x}/k_\gamma \right)\in\left(0,\pi\right)$ ($\mathbf{k}_\gamma$ is the photon wave-vector) and are symmetrized relative to $\theta_\gamma=0$. The laser-to-photon energy conversion efficiency,  $\eta_\gamma \equiv \mathcal{E}_\gamma/\mathcal{E}_L$, is indicated in each case.}
\label{fig:uc_plasmas_spectra}
\end{figure}

Interestingly, the $p_x$ profile of the return current presents anharmonic oscillations at $\lambda_{\rm mod} \approx 1.5\lambda_L$, resulting in strong density modulations ($\Delta n_e/n_e \gtrsim 1$) inside the laser pulse. The related maxima in $\vert p_x \vert$, when coinciding with $B_z$ extrema, yield peaks in the radiated power density profile (translating into $\sim 5\,\rm fs$ time scale fluctuations in the spatially averaged radiated power, not shown here). This is expected as those high-energy counterstreaming electrons are those optimizing the quantum parameter $\chi_e \simeq \gamma (1-v_x/c)a_L \hbar \omega_L/m_e c^2 \simeq 2 \gamma a_L \hbar \omega_L/m_e c^2$ (for purely counterstreaming electrons of typical energy $\gamma$ and longitudinal velocity $v_x$), resulting in a backward-directed radiated power (per electron) $P_{\rm cl} \simeq (2/3) \alpha_f  m_e c^2 \chi_e^2/\tau_C \simeq (8/3) (r_e \omega_L/c) \omega_L m_e c^2 (\gamma a_L)^2$, with $r_e$ being the classical electron radius, and assuming negligible quantum corrections \cite{PPCFKirk2009}. The large $\vert p_y \vert$ momenta of the counterstreaming electrons at the emission peaks account for the extended backward-directed emission lobe seen in the angular spectrum plotted (as a blue curve) in Fig.~\ref{fig:uc_plasmas_spectra}. A weaker and narrower forward-directed component is also visible, due to the reflected part of the laser wave being scattered by the forward-moving electrons. The total laser-to-photon energy conversion, defined as the fraction of the injected laser energy radiated into $\geq 10 \,\rm keV$ energy photons, is measured to be $\eta_\gamma \simeq 13\,\%$ at the end of the simulation ($t=150\,\rm fs$).

The above emission scenario, hinging on the electrons injected back into the electromagnetic wave at the laser front, was first investigated in Refs.~\onlinecite{PRLBrady2012, PoPBrady2014}, where it was termed re-injected electron synchrotron emission (RESE), and found to yield the largest radiation yield at $I_L \gtrsim 10^{22}\,\rm Wcm^{-2}$. The overall description provided in those works is consistent with our results, except regarding the quantitative estimate of the radiation burst time scale ($\sim 5\,\rm fs$ here).
This time was interpreted as that needed for the compressed electrons at the laser front to build up an electrostatic field ($E_x \simeq en_{e0}ct/\epsilon_0$) exceeding the $\mathbf{v}\times \mathbf{B} \propto a_L$ force, thus reflecting them toward the laser source. This reasoning yields a `breakdown time'\cite{PoPBrady2014}, $\tau_{bd} \simeq a_L (n_c/n_{e0}) \omega_L^{-1}$. Under the present conditions, we should have $c\tau_{bd} \simeq 0.8\,\rm \mu m$, which is about half the observed spacing of the $P_\gamma$ peaks, $\lambda_{\rm mod}=1.5\,\rm \mu m$.

Rather, we propose the following simple explanation for the modulations affecting the $p_x <0$ hot electrons. Let us consider their motion in the rest frame of the laser front, in which the Doppler-shifted laser frequency is $\omega_L' = \omega_L \sqrt{(1-v_f/c)/(1+v_f/c)}$ (assuming $k_L \approx \omega_L/c$). The electrons impinging on the laser front from the unperturbed plasma experience the $2\omega_L'$-oscillating component of the laser's ponderomotive force while being injected downstream at $v_x \approx -c$. As a consequence, a current modulation is induced with wavenumber $k'_{\rm mod} = -2\omega_L'/c$. In the laboratory frame, this wavenumber becomes $k_{\rm mod} = - 2\gamma_f  (1-v_f/c)\omega'_L/c = - 2 [(1-v_f/c)/(1+v_f/c)] \omega_L/c$, corresponding to a wavelength
\begin{equation}
  \lambda_{\rm mod} = [(1+v_f/c)/(1-v_f/c)]\lambda_L/2 \,.  \label{eq:modulation_counterstreaming}
\end{equation}
In the present case, where $v_f/c \simeq 0.47$, one expects $\lambda_{\rm mod} \approx 1.4\,\rm \mu m$, in good agreement with the simulation.

The observation that the radiation is mainly backward directed and emitted as bursts throughout the irradiated region allows for a rough estimate of the total radiation yield,
\begin{equation}
\eta_\gamma = \xi \frac{P_{\rm cl} n_{h<} v_f t }{I_L} \,,
\end{equation}
where we have introduced $n_{h<}$ the density of the counterstreaming ($p_x<0$) electrons, and $\xi$ the ratio of the burst length to its spacing $\lambda_{\rm mod}$. Further assuming a mean electron energy $\langle \gamma \rangle \approx a_L$ -- fairly consistent with Fig.~\ref{fig:uc_plasmas_interaction}(b) -- gives 
\begin{equation}
  \eta_\gamma \approx  \kappa \xi \frac{n_{h<}}{n_c} \frac{v_f t \omega_L}{c} a_L^2 \,, \label{eq:sync_rese}
\end{equation}
with $\kappa \equiv (16/3) (r_e \omega_L/c) \simeq 9.44\times 10^{-8}$. Taking $n_{h<} = n_{e0}/2$ and $\xi = 0.1$ leads to $\eta_\gamma \approx 8\,\%$ at $t=150\,\rm fs$. This value is comparable with the simulation value $\eta_\gamma \simeq 13\,\%$. The difference is attributed to uncertainties in the estimation of the electron parameters, to modulations in the laser field strength, and to the neglect of the forward-directed radiation (due to the forward-moving electrons interacting with the light reflected off the laser front).

For completeness, we have repeated the same simulation with a Gaussian laser pulse of $30\,\rm fs$ FWHM duration, impinging onto either a semi-infinite or $1\,\rm \mu m$ thick $\rm C^{6+}$ plasma of electron density $n_{e0}=17n_c$.
As expected, the semi-infinite plasma yields a spatially averaged radiated power at the pulse maximum ($d\mathcal{E}_\gamma/dt \simeq 0.045\,\rm J\,fs^{-1}\,\mu m^{-1}$) close to that measured at the same time with a constant laser drive. It also leads to a similar radiated angular spectrum (compare the blue and green curves in Fig.~\ref{fig:uc_plasmas_spectra}), although with a more pronounced transverse component ($\theta_\gamma \simeq \pi/2$). This change is ascribed to the energy depletion of the short laser pulse as it propagates through the plasma, which leads to near-transparency interaction conditions and favors transverse emission, as observed previously~\cite{PoPChang2017, PPCFMartinez2018}. Our simulation also predicts that due to progressive depletion of the laser pulse, the total radiated power starts diminishing after $t\simeq 50\,\rm fs$, and falls below $10\,\%$ of its maximum value at $t \gtrsim 100\,\rm fs$ (not shown).

\begin{figure*}[t]
\centering
\includegraphics[width=0.99\textwidth]{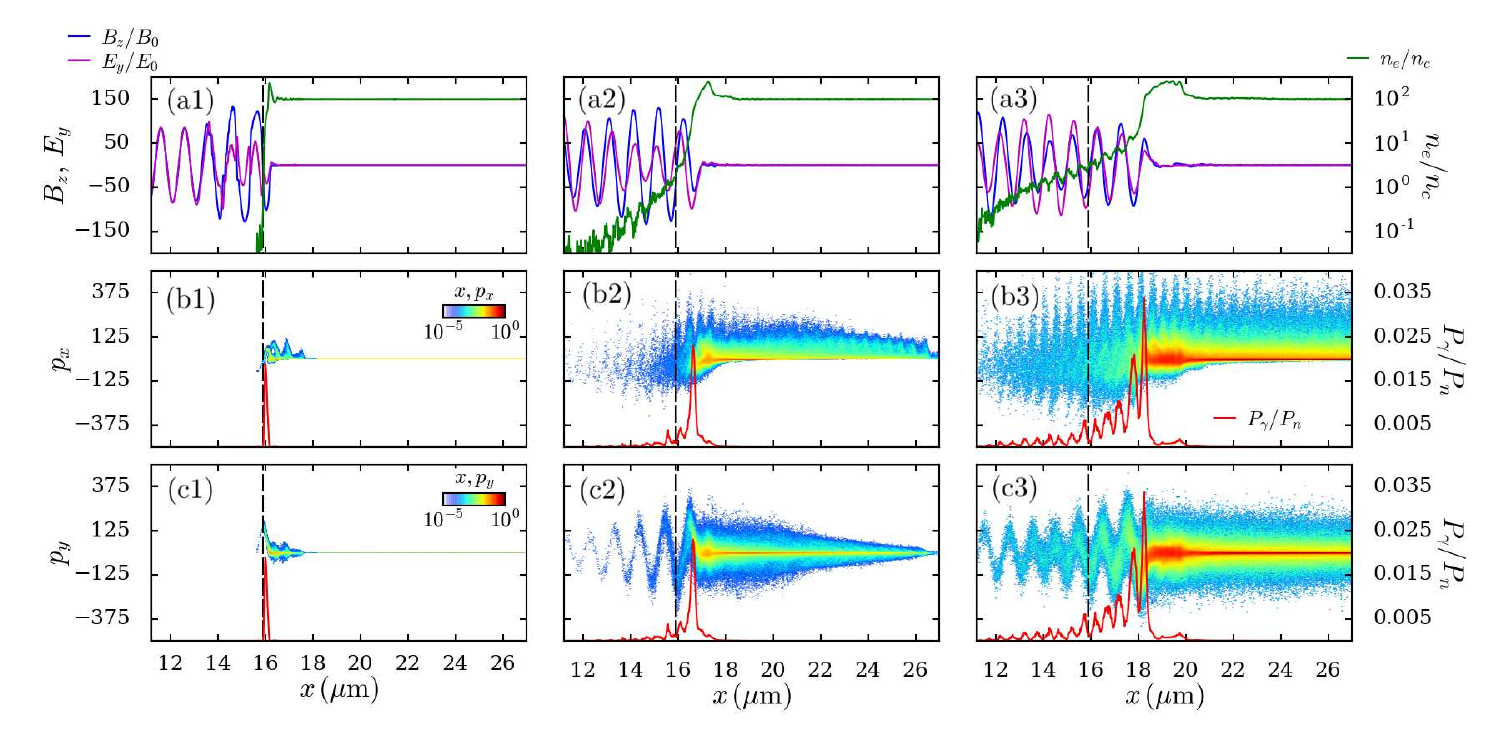}
\caption{Interaction of a semi-infinite, electromagnetic plane wave ($I_L=10^{22}\,\rm Wcm^{-2}$) with a semi-infinite, relativistically overcritical C$^{6+}$ plasma ($n_{e0}=100n_c$). (a1-a3) Longitudinal lineouts of the $E_y$ (purple) and $B_z$ (blue) field components and of the electron density $n_e$ (green). (b1-b3) $x-p_x$ electron phase space (averaged along $y$) (red). (c1-c3) $x-p_y$ electron phase space (averaged along $y$). In (b1-b3) and (c1-c3), the red curve plots a lineout of the synchrotron radiated power density, $P_\gamma$ (red). The three columns correspond to the interaction times:  (a1-c1) $t=4\,\rm fs$, (a2-c2) $36\,\rm fs$ and (a3-c3) $100\,\rm fs$.}
\label{fig:oc_plasmas_interaction}
\end{figure*} 

When considering a finite ($1\,\rm \mu m$) plasma thickness, the radiated power is reduced by approximately an order of magnitude ($\eta_\gamma \simeq 1\,\%$), and the emission is more concentrated to the backward direction (Fig.~\ref{fig:uc_plasmas_spectra}). There are two main reasons for these features. The first is that, unlike what occurs in a semi-infinite plasma, where the counterstreaming electrons that mainly account for high-energy radiation are continually replenished at the laser front (as long as the laser has not been strongly depleted), these are now electrostatically confined around the target, so that the radiation only occurs during the transit time of the laser pulse. Second, because of the short interaction time and the rapid plasma expansion, there is no significant laser reflection; this reduces the radiative contribution of the high-energy $p_x>0$ electrons, and explains the vanishing forward emission.

\subsection{Relativistically overcritical plasma}
\label{subsec:overcritical}

We now address the case of a semi-infinite, relativistically overcritical plasma ($n_{e0}/n_c=100$) illuminated by a semi-infinite, $10^{22}\,\rm Wcm^{-2}$ intensity laser wave. Figures~\ref{fig:oc_plasmas_interaction}(a1-c3) present the main features of the interaction at three successive times.

\begin{figure}[!t]
\centering
\includegraphics[width=0.48\textwidth]{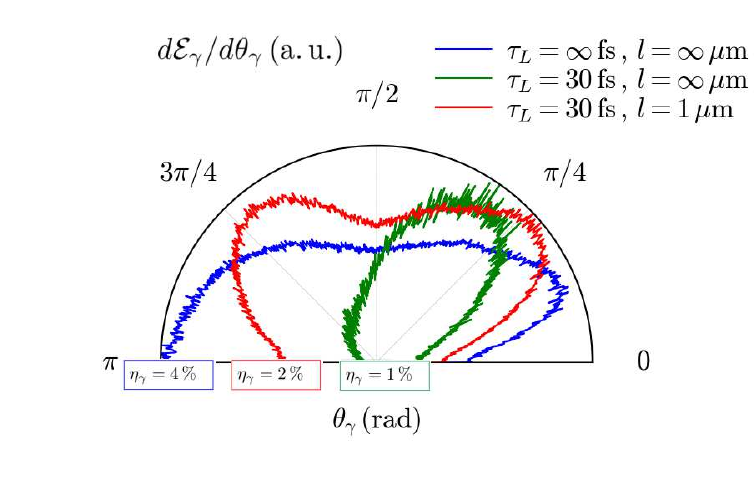}
\caption{Same as Fig.~\ref{fig:uc_plasmas_spectra} for a C$^{6+}$ plasma with $n_e = 100n_c$.
}
\label{fig:oc_plasmas_spectra}
\end{figure}

The front-side electrons are energized through vacuum/$J\times B$ heating~\cite{PoPBauer2007, PREMay2011, PoPDebayle2013}, leading to periodic injection of fast electron bunches into the plasma at twice the laser frequency. The $x-p_x$ and $x-p_y$ electron phase spaces of Figs.~\ref{fig:oc_plasmas_interaction}(b1-c1) capture the instant ($t=4\,\rm fs$) when the skin-layer electrons accelerated by the $E_y$ [purple curve in Fig.~\ref{fig:oc_plasmas_interaction}(a1)] component of the standing wave set up in vacuum (near the plasma boundary) have acquired their maximum transverse momenta, and are being rotated by the $B_z$ field [blue curve in Fig.~\ref{fig:oc_plasmas_interaction}(a1)] toward the plasma~\cite{PREMay2011}. The radiated power density [red curve in Figs.~\ref{fig:oc_plasmas_interaction}(b1-c1)] peaks just in front of the steep plasma boundary, where $B_z$ is at its highest, and the accelerated electrons are characterized by $p_x \approx 120 m_ec > \vert p_y \vert \approx 50m_ec$. This gives rise to a forward/oblique emission lobe extending from $\theta_\gamma \simeq 30$ to $90^\circ$, as seen in the angular spectrum of Fig.~\ref{fig:oc_plasmas_spectra} (blue curve).
Note that the distorted $E_y$ and $B_z$ field profiles in vacuum [Fig.~\ref{fig:oc_plasmas_interaction}(a1)] are due to high-order harmonic generation from the oscillating plasma surface~\cite{PREGonoskov2011}.

At a later time ($t=36\,\rm fs$), the plasma temperature has strongly increased, and the plasma boundary, pushed by the laser's radiation pressure, has developed both a bump and a longer scale-length density profile [see Fig.~\ref{fig:oc_plasmas_interaction}(a2)]. The density bump is the signature of an electrostatic shock \cite{PRLSilva2004}, which traps part of the fast electrons behind the laser ``piston'' (see the electrons with $p_x<0$ around $x \simeq 17-18\,\rm \mu m$).
The expanding dilute portion of the density profile ($x< 16\,\rm \mu m$) corresponds to the few electrons leaked through the ponderomotive barrier at the plasma boundary and moving across the standing wave.
Synchrotron radiation then mainly occurs within an enlarged ($\sim 0.3\lambda_L$ thick) region that encompasses the skin layer and the lower-density electron cloud in front of it.
The radiated power density culminates around $\sim 20-40 n_c$ electron densities, where the electron phase space shows high positive and negative $p_x$ values with, typically, $\vert p_x\vert \sim p_y$. The resulting synchrotron emission is thus spread over a broad angular range in both forward and backward directions.

At an even later stage ($t=100\,\rm fs$), a larger number of electrons have escaped into the vacuum,
forming an extended, relativistically undercritical shelf modulated at $\lambda_L/2$ [Fig.~\ref{fig:oc_plasmas_interaction}(a3)]. There, the high-energy electrons exhibit an approximately even $p_x$ momentum distribution, and mainly radiate around the $E_y$ (or $B_z$) extrema of the laser wave. Due to laser absorption, the radiation then mostly arises from the electrons counterstreaming against the incoming laser wave, hence accounting for the backward-directed ($\theta_\gamma \simeq \pi$) lobe visible in
Fig.~\ref{fig:oc_plasmas_spectra}. At the end of the simulation ($t=150\,\rm fs$), about $4\,\%$ of the incident light energy is radiated away, which is about three times less than at $n_{e0}=17n_c$. 

If the plasma is irradiated by a $30\,\rm fs$ laser pulse, the synchrotron efficiency drops to $\eta_\gamma \simeq 1\,\%$ and, due to the shortened interaction time, the above-discussed, late-time backward components of the synchrotron emission vanishes (green curve in Fig.~\ref{fig:oc_plasmas_spectra}). Changing to a $1\,\rm \mu m$ foil target (while keeping the same laser parameters) improves the radiation efficiency ($\eta_\gamma \simeq 2\,\%$), unlike in the $n_{e0}=17n_c$ case. This differing trend stems from the fact that in the latter transparent regime the radiation occurs volumetrically, and so the radiation yield decreases in thinner targets. At $n_{e0}=100n_c$, by contrast, the foil remains opaque throughout the interaction: the emission is confined to the  front side (precisely, in the $\sim 5-100n_c$ density plasma shelf preceding the laser-compressed skin layer), and its efficiency increases when a larger number of high-energy electrons propagate against the laser wave, as happens due to electrostatic reflection at the target backside. Consequently, the radiation from the $1\,\rm \mu m$ foil exhibits two forward and backward-directed lobes, about symmetric relative to the transverse axis (red curve in Fig.~\ref{fig:oc_plasmas_spectra}).

\section{Competition between Bremsstrahlung and synchrotron emission in copper foil targets}
\label{sec:comp_brem_sync}

We now study the relative contributions of Bremsstrahlung and synchrotron emission to the total high-energy radiation from a laser-driven thin solid foil. In contrast to the few previous studies on this subject \cite{PoPPandit2012, EPJDWan2017, PPCFVyskocil2018}, which essentially focused on the laser intensity dependence of those two radiative processes, we will consider fixed laser parameters and a single target material (Cu), and will investigate, through 2D simulations, the influence of the target thickness, varied from a few nm to a few $\rm \mu m$.

\subsection{Numerical setup and modeling}

The 2-D simulations reported below consider a laser pulse propagating in the $+x$ direction, linearly polarized along $y$, with a wavelength $\lambda_L=1\,\rm \mu m$
and a maximum intensity $I_L=10^{22}\,\rm W cm^{-2}$ ($a_L =85$). Moreover, it has a Gaussian temporal profile of $50\,\rm fs$ FWHM and a Gaussian transverse profile of $5\,\rm \mu m$ FWHM.
The target consists of a solid-density copper plasma slab of thickness $16\,\mathrm{nm} \le l \le 5\,\rm \mu m$. It is initialized with a $200\,\rm eV$ temperature and a $Z^*=25$ ionization state, corresponding to an electron density $n_{e0}\simeq 2000 n_c$. Its front and rear sides are coated with $3.2\,\rm nm$ thick hydrogen layers of atomic density $n_\mathrm{H}=50n_c$, which model the hydrogen-rich surface contaminants usually responsible for proton beam generation in laser experiments~\cite{RMPMacchi2013}. Note that an ultrahigh intensity contrast is implicitly assumed; otherwise, the front-side hydrogen layer is expected to be blown away by the laser prepulse.

The domain dimensions are $L_x \times L_y = 127 \times 40\,\rm \mu m^2$ with a mesh size $\Delta x = \Delta y = 3.2\,\rm nm$
The number of macro-particles per cell and species is adjusted depending on the foil thickness to limit the numerical cost. Specifically, it is
varied from $2000$ to $375$ for $l \in (16,32,51,100)\,\rm nm$, and from 40 to 10 for $l \in (0.5,1,5)\,\rm \mu m$. Absorbing boundary conditions for particles and fields are employed in both $x$ and $y$ directions. The simulations are run over durations ranging from $270\,\rm fs$ ($l=16\,\rm nm$) to $\approx 800\,\rm fs$ ($l=5\,\rm \mu m$).

Besides Bremsstrahlung and synchrotron emission, all simulations self-consistently describe elastic Coulomb collisions as well as impact and field induced ionization\cite{PoPNuter2011, PoPPerez2012}. Bremsstrahlung is modeled using the Monte Carlo scheme developed in Ref.~\onlinecite{PoPMartinez2019}, taking account of both Thomas-Fermi and Debye-type screening effects. 

\subsection{Target thickness dependence of the radiation yield}

\begin{figure}[tbh!]
\centering
\includegraphics[width=0.48\textwidth]{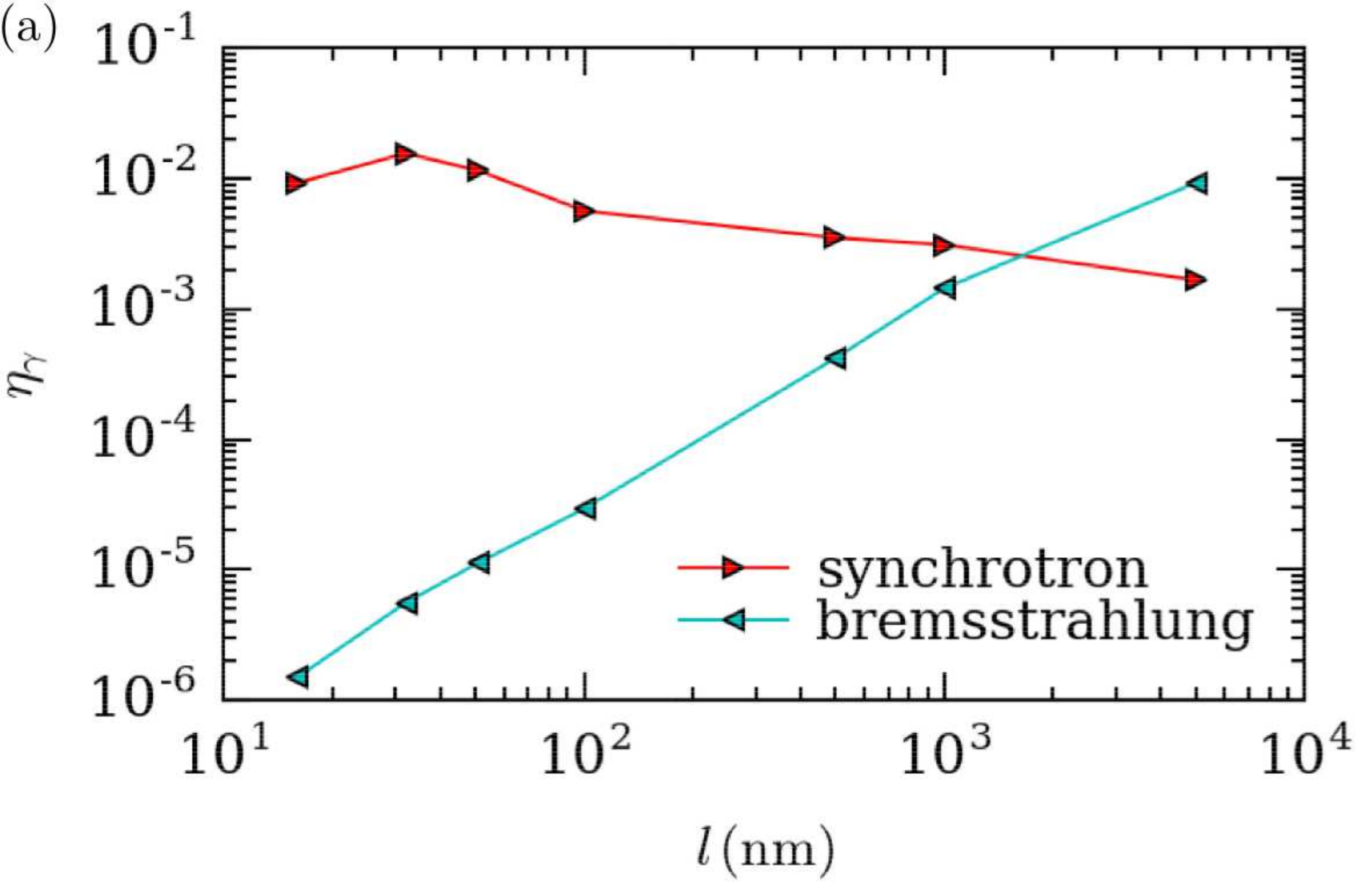}
\includegraphics[width=0.48\textwidth]{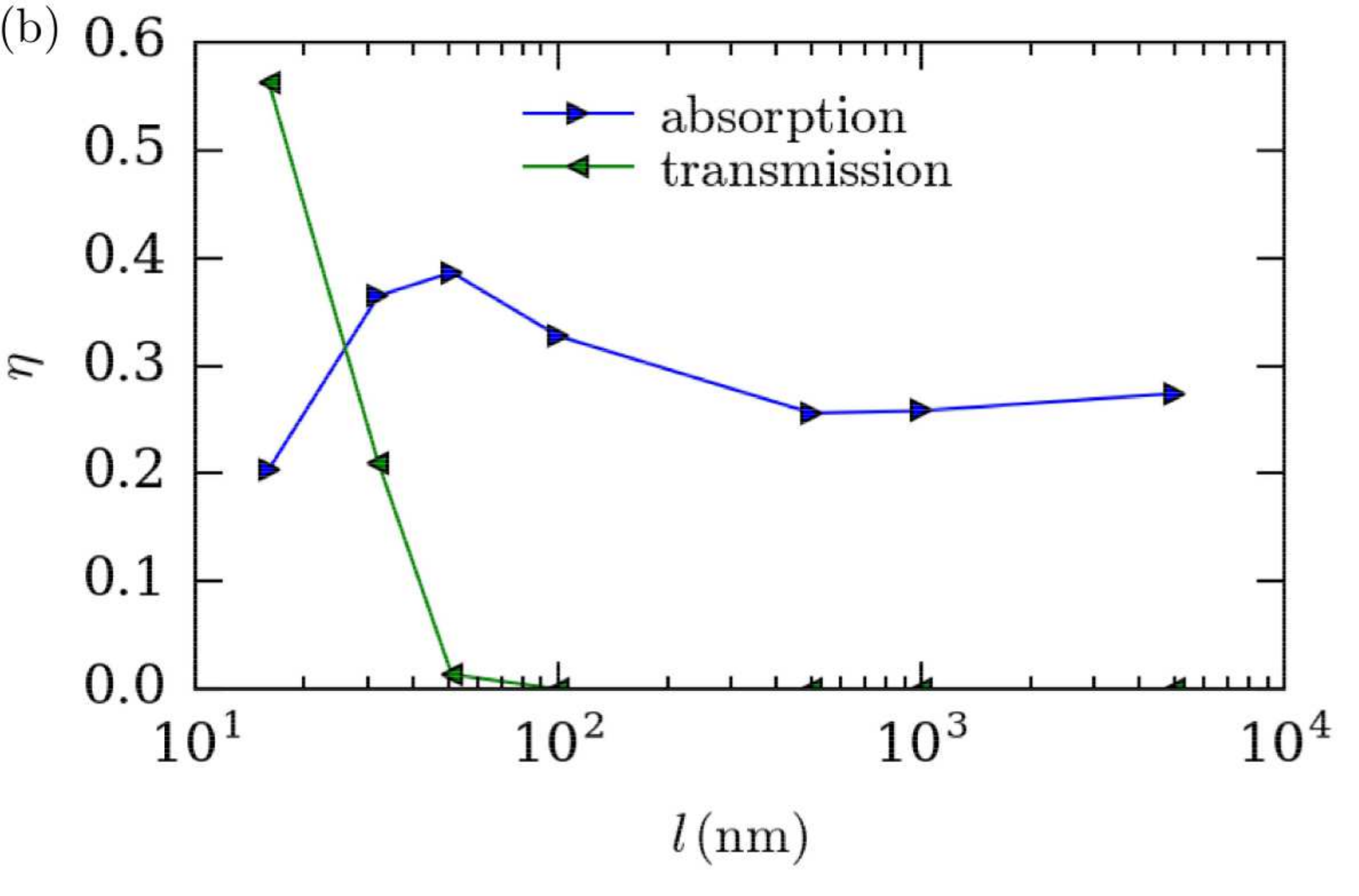}
\caption{2D PIC simulations of an ultraintense ($10^{22}\,\rm W\,cm^{-2}$), ultrashort ($50\,\rm fs$) and tightly focused ($5\,\rm \mu m$) laser pulse interacting with a copper foil target. (a) Synchrotron (red triangles) and Bremsstrahlung (cyan triangles) conversion efficiencies into $>10\,\rm keV$ photons as a function of the target thickness. (b) Laser absorption (blue triangles) and transmission (green triangles) as a function of the target thickness.}
\label{fig:eff}
\end{figure}

The energy conversion efficiencies ($\eta_\gamma$) of Bremsstrahlung (cyan) and synchrotron emission (red) into $>10\,\rm keV$ photons are plotted in Fig.~\ref{fig:eff}(a) as a function of the target thickness. The synchrotron efficiency initially grows from $\eta_\gamma = 0.8\,\%$ at $l=16\,\rm nm$ to a maximum of $1.5\,\%$ at $l=32\,\rm nm$.
It slowly decreases at larger thicknesses, reaching $0.2\,\%$ at $l=5\,\rm \mu m$. By comparison, the Bremsstrahlung efficiency steadily rises with thicker targets, scaling as $\eta_\gamma \propto l^{1.5}$ in the thickness range considered. Specifically, it increases from $\eta_\gamma = 2\times 10^{-4}\,\%$ at $l=16\,\rm nm$ to $1\,\%$ at $l=5\,\rm \mu m$. An important finding is that the Bremsstrahlung and synchrotron curves cross each other for $l\simeq 1-2\,\rm \mu m$, in which case they both attain $\eta_\gamma \simeq 0.3\,\%$. 

It is interesting to confront those results with the corresponding variations in the laser absorption and transmission rates, displayed in Fig.~\ref{fig:eff}(b). The twofold increase in the absorption coefficient between $l=16\,\rm$ and $32\,\rm nm$ is similar to that observed in the synchrotron efficiency. The absorption culminates in a plateau around $l=32-50\,\rm nm$, which also encompasses the maxima of the synchrotron efficiency. The transmission coefficient reaches $\sim 56\,\%$ at $l=16\,\rm nm$ and abruptly drops within the plateau (to $\sim 20\%$ at $l=32\,\rm nm$ and $\sim 0.1\,\%$ at $l=50\,\rm nm$). In light of the results of Sec.~\ref{sec:sync_unif_plasmas}, the slightly better synchrotron performance at $l=32\,\rm nm$ than at $l=50\,\rm nm$ is ascribed to the partial transparency of the target, which allows the electrons to experience the full strength of the laser fields.

The conditions of strong laser absorption and significant transmission that maximize synchrotron emission are also known to enhance ion acceleration from thin foils driven by femtosecond laser pulses \cite{PoPdHumieres2005, PRLEsirkepov2006, PRSTABBrantov2015, ArxivFerri2020}. The optimum thickness for ion acceleration has been found to be $l_{\rm ion} \simeq 0.5 \lambda_L a_L (n_c/n_{e0})$ \cite{PRLEsirkepov2006, PRSTABBrantov2015}, close to the threshold thickness for self-induced relativistic transparency \cite{PoPVshivkov1998}. In the present case ($a_L =85$, $n_{e0}/n_c=2000$), one has $l_{\rm ion} \simeq 21\,\rm nm$, a bit lower than the synchrotron-optimizing thickness $l\simeq 32\,\rm nm$.

At larger thickness ($l>50\,\rm nm$), our simulations predict that the absorption coefficient first decreases before stagnating at $\sim 25\,\%$ for $l\ge 0.5\,\rm \mu m$. This mere $\sim 30\,\%$ decrease in the laser absorption is accompanied by a more pronounced (by an order of magnitude) drop in the synchrotron efficiency. This further shows that the laser absorption is not the only figure-of-merit for ensuring strong synchrotron emission. 

\subsection{Illustrative cases}
\label{subsec:illustrative_cases}

\subsubsection{Transparent 32-nm-thick target}
\label{subsubsec:32nm_target}

\begin{figure}[t]
\centering
\includegraphics[width=0.45\textwidth]{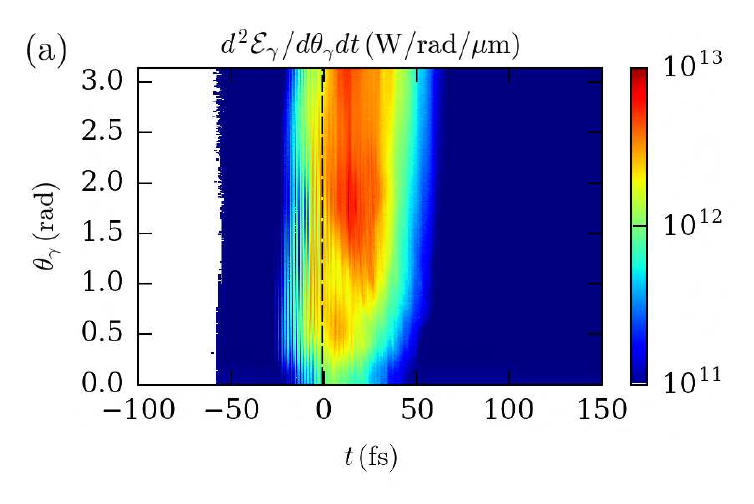}
\includegraphics[width=0.45\textwidth]{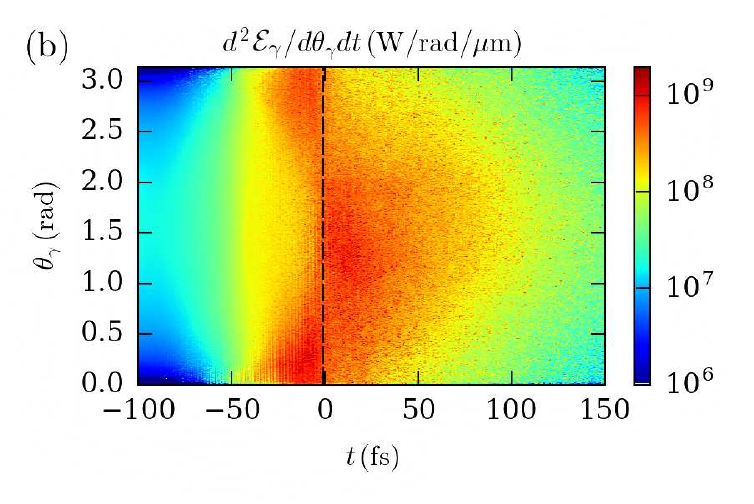}
\caption{Time evolution of the angle-resolved (a) synchrotron and (b) Bremsstrahlung power spectrum from a 32-nm-thick Cu foil target. The dashed line at $t\simeq 0\,\rm fs$ indicates both the laser pulse maximum and the onset of relativistic transparency.
}
\label{fig:dPdth_l32nm_2D}
\end{figure}

We now focus on the radiation dynamics in the $l=32\,\rm \mu m$ foil that maximizes the synchrotron efficiency. Figure~\ref{fig:dPdth_l32nm_2D}(a) displays the time evolution of the synchrotron angular spectrum. A transition is seen to occur around the on-target laser peak ($t=0\,\rm fs$), which also coincides with the onset of relativistic transparency. Before transparency occurs, the Cu bulk plasma is compressed by the radiation pressure and set into motion as a whole -- a process known as light-sail-type acceleration \cite{HPLSEMacchi2014}. Some of the fast electrons (accelerated up to $\vert p_x\vert \simeq 100\,m_ec$) recirculating around the bulk plasma are capable of passing through the laser piston to form a relativistically under/near-critical shelf in front of the compressed boundary. As in the scenario considered in Sec.~\ref{subsec:overcritical}, synchrotron emission then mainly takes place in this relatively dilute ($10-100\,n_c$) expanding cloud. Due to significant laser reflection, both the forward and backward-moving high-energy electrons contribute to the radiation, the spectrum of which thus presents broad emission lobes in the forward and backward directions.

\begin{figure}[tb]
\centering
\includegraphics[width=0.45\textwidth]{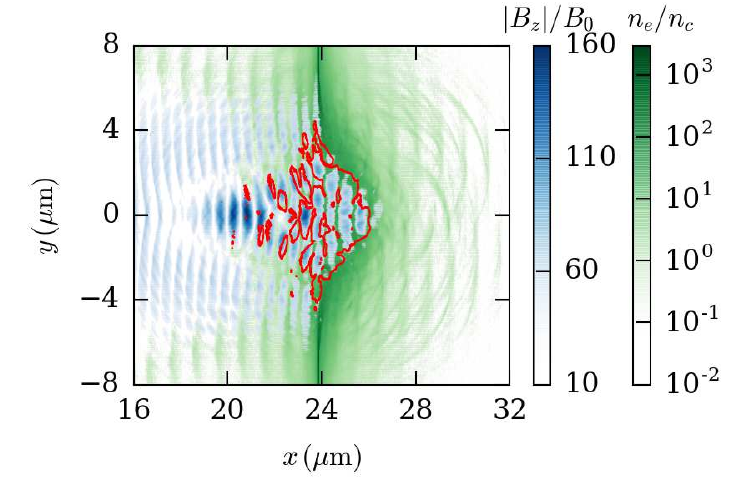}
\caption{32-nm-thick copper foil target: Spatial distributions of the modulus of the magnetic field $\vert B_z \vert$ (blue colormap) and the electron density $n_e$ (blue colormap) at $t=12\,\rm fs$ after the on-target laser peak. The closed red curves are isocontour lines (at $P_\gamma = 10^{23}\,\rm Wcm^{-3}$) of the synchrotron radiated power density.}
\label{fig:bz_ne_pr_l32nm_2D_t12fs_it30000}
\end{figure}

Figure~\ref{fig:bz_ne_pr_l32nm_2D_t12fs_it30000} illustrates the laser-plasma interaction and the emissive region at $t=12\,\rm fs$, just after the target has turned transparent to the laser light. This instant is when the (spatially integrated) synchrotron power is at its highest. The pseudocolor maps show the spatial distributions of the absolute magnetic field strength (blue) and of the electron density (green). Overlaid is an isocountour (at $P_\gamma = 10^{23}\,\rm W\,cm^{-3}$) of the instantaneous synchrotron power density (red). Rayleigh-Taylor-like modulations with spatial scale $\sim \lambda_L$ have developed in the irradiated region, breaking the translational invariance along $y$, and hence enhancing the electron energization (above $\vert p_x \vert = 400\,m_ec$)~\cite{PREMay2011}. This disrupts the early-time balance between the radiation and particle momentum fluxes~\cite{PRLYan2008}, and leads to the accelerated Cu plasma being bored through by the laser pulse. Close inspection reveals that synchrotron emission is then concentrated in the laser-filled bulk plasma turned undercritical, of $\sim 3\,\rm \mu m$ length and $3-40\,n_c$ electron density. The time-resolved synchrotron energy spectrum, which was nearly isotropic early in the interaction, then increases in intensity and becomes mainly backward directed (at angles $\theta_\gamma \gtrsim \pi/2$). As the laser pulse traverses the plasma, the average laser field strength experienced by the (electrostatically confined) relativistic electrons diminishes and so does the synchrotron power, which scales as $\sim \langle \gamma(t) \rangle  a_L(t)^2$. The synchrotron emission becomes negligible once the laser pulse has travelled past the plasma ($t \simeq 50 \,\rm fs$).

Figure~\ref{fig:dPdth_l32nm_2D}(b) shows the Bremsstrahlung angular power spectrum as a function of time. Overall, its maximum values are about 4 orders of magnitude lower than those of the synchrotron spectrum [note the different scales of the colorbars in Figs.~\ref{fig:dPdth_l32nm_2D}(a) and (b)]. 
The Bremsstrahlung emission presents a strong rise at $t \simeq -50\,\rm fs$ (when the on-target laser intensity reaches $6.5\times 10^{20}\,\rm Wcm^{-2}$), both in the forward ($\theta_\gamma =0$) and the backward ($\theta_\gamma=\pi$) directions. This longitudinal emission is due to the high-energy electrons recirculating across the dense, high-$Z$ Cu layer. At early times, the Cu layer is still planar and opaque to the laser field; therefore, owing to the quasi-1D interaction geometry, the transverse momenta of the fast electrons vanish inside the target, leading to peaked forward and backward emission lobes (recall that the Bremsstrahlung photons are emitted within an angular cone of $\sim \gamma^{-1}$ aperture along the electron direction). At later times, however, the angular distribution of the hot electrons broadens as a result of transverse surface modulations, and hence the angular Bremsstrahlung spectrum becomes increasingly isotropic.

Similarly to synchrotron radiation, the Bremsstrahlung radiated power culminates at the laser peak, yet decays away over a longer time scale ($\sim 100\,\rm fs$ vs $\sim 50\,\rm fs$ for synchrotron) in the subsequent transparency regime. This decreasing trend can be understood from the following approximate expression of the (space-integrated) Bremsstrahlung power, valid in the ultra-relativistic limit and neglecting electron screening \cite{APJQuigg1968}:
\begin{align} \label{eq:Brem_power}
  \frac{dE_{b,h}}{dt} &= 12\alpha r_e^2 Z^2 m_e c^3 D_h l_{\rm Cu} \langle n_{\rm Cu} \rangle \langle n_h \rangle \langle \gamma_h \rangle \nonumber \\
  &\times \left[ \log \left(2 \langle \gamma_h \rangle\right) +  0.92 \right] \,,
\end{align}
where $l_{\rm Cu}$ denotes the longitudinal width of the expanding bulk copper plasma, with mean ion density $\langle n_{\rm Cu} \rangle$. We have also introduced $D_h$ the transverse width of the hot-electron cloud,  with mean density $\langle n_h \rangle$ and energy $\langle \gamma_h \rangle$. Note that the mean energy of Bremsstrahlung photons is $\langle \hbar \omega \rangle \approx m_e c^2 \langle \gamma_h \rangle/3$ \cite{bookDermer2009}. Since the areal density $\langle n_{\rm Cu} \rangle l_{\rm Cu}$ is approximately constant, the Bremsstrahlung power should vary as $dE_{b,h}/dt \propto D_h \langle n_h \rangle \langle \gamma_h \rangle (t)$. 
Now, as the target expands and becomes increasingly quasineutral, most of the hot electrons are confined within the Cu bulk plasma, so that their longitudinal extent approximately coincides with $l_{\rm Cu}$.
Introducing the total hot electron energy $E_h(t) \simeq l_{\rm Cu} D_h \langle n_h \rangle \langle \gamma_h \rangle (t)$, and noting that  $D_h l_{\rm Cu} \langle n_h \rangle \simeq \rm cst$, one obtains $dE_{b,h}/dt \propto E_h(t)/l_{\rm Cu}(t)$. 

By looking at the dynamics of the target expansion and of the particle kinetic energies, we have checked that the above scaling is consistent with the observed evolution of the Bremsstrahlung power following the laser pulse maximum (and the onset of the target transparency). Over the timespan $0< t < 100\,\rm fs$, the spatial extent of the bulk Cu plasma varies by a factor of $\sim 1.5$ while, owing to energy transfer to Cu ions (which carry about $30\,\%$ of the laser energy at $t=100\,\rm fs$), the total electron energy drops by a factor of $\sim 5$, so that $dE_{b,h}/dt$ should decay by a factor of $\sim 7.5$. This prediction reasonably agrees with the then measured 5-fold decrease in the Bremsstrahlung power [calculated from integration over $\theta_\gamma$ of the spectrum shown Fig.~\ref{fig:dPdth_l32nm_2D}(b)], which goes from $1.6 \times 10^9\,\rm W \,\mu m^{-1}$ to $3.1\times 10^8\,\rm W\,\mu m^{-1}$.

\subsubsection{Opaque 5~µm-thick target}
\label{subsubsec:5mum_target}

\begin{figure}[t]
\centering
\includegraphics[width=0.45\textwidth]{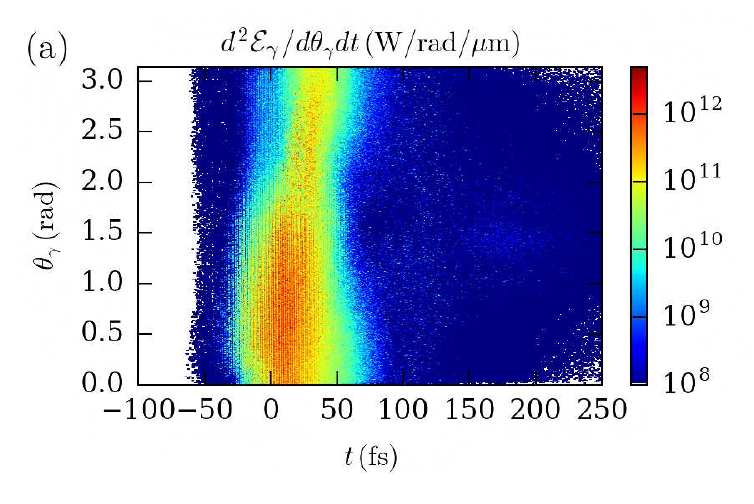}
\includegraphics[width=0.45\textwidth]{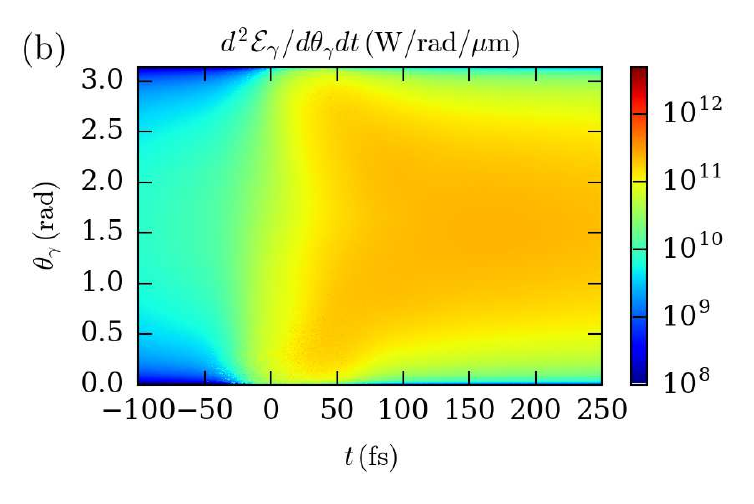}
\caption{Same as Fig.~\ref{fig:dPdth_l32nm_2D} for the $5\,\rm \mu m$ thick copper foil target.}
\label{fig:dPdth_l5mum_2D}
\end{figure}

We now consider the radiation from a $5\,\rm \mu m$ Cu foil, that is, the thickest target considered in our 2D simulation study. This target remains opaque to the laser light throughout the interaction. At the final simulation time ($t=250\,\rm fs$), while expanding at its rear and -- to a lower extent -- front sides, the thickness of the solid-density Cu layer is still about $4/5$ of its initial value. 

\begin{figure}[tb]
\centering
\includegraphics[width=0.45\textwidth]{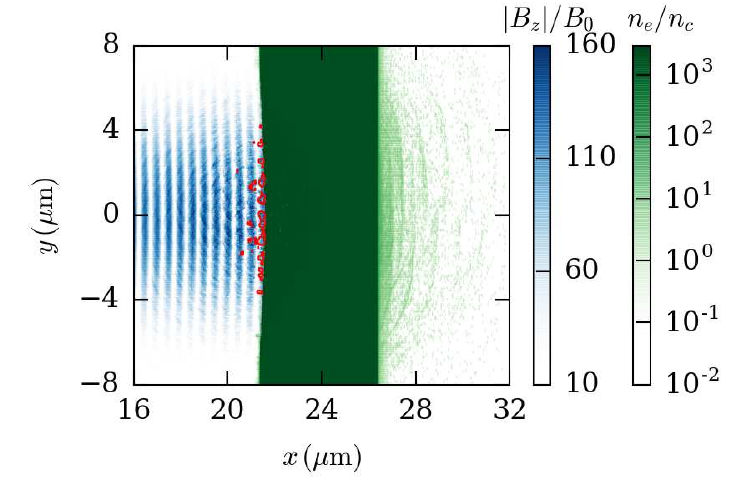}
\caption{Same as Fig.~\ref{fig:bz_ne_pr_l32nm_2D_t12fs_it30000} for a $5\,\rm \mu m$ thick copper foil target.}  
\label{fig:bz_ne_pr_l5mum_2D_t07fs_it2800}.
\end{figure}

\begin{figure}[tb]
\centering
\includegraphics[width=0.45\textwidth]{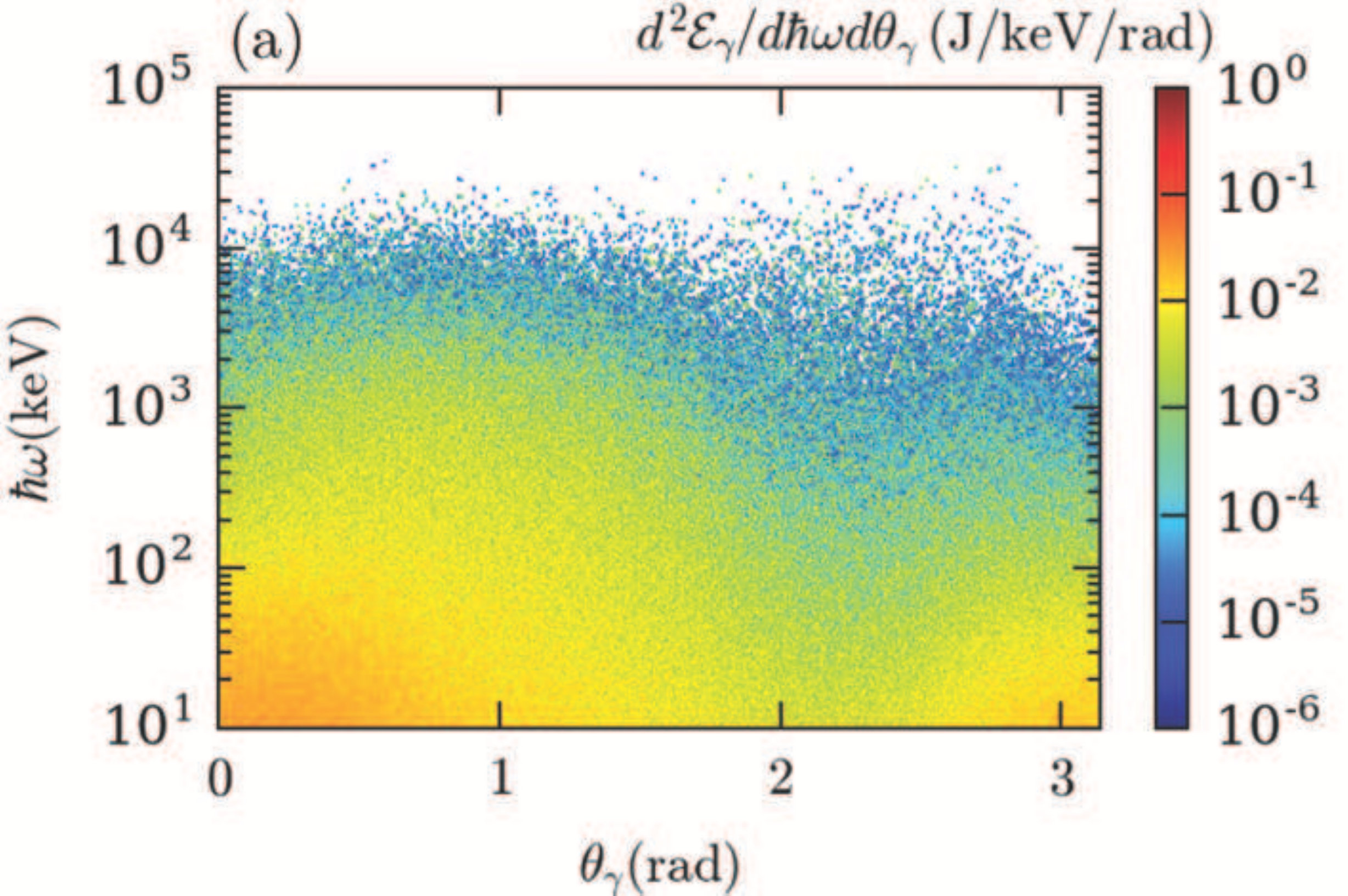}
\includegraphics[width=0.45\textwidth]{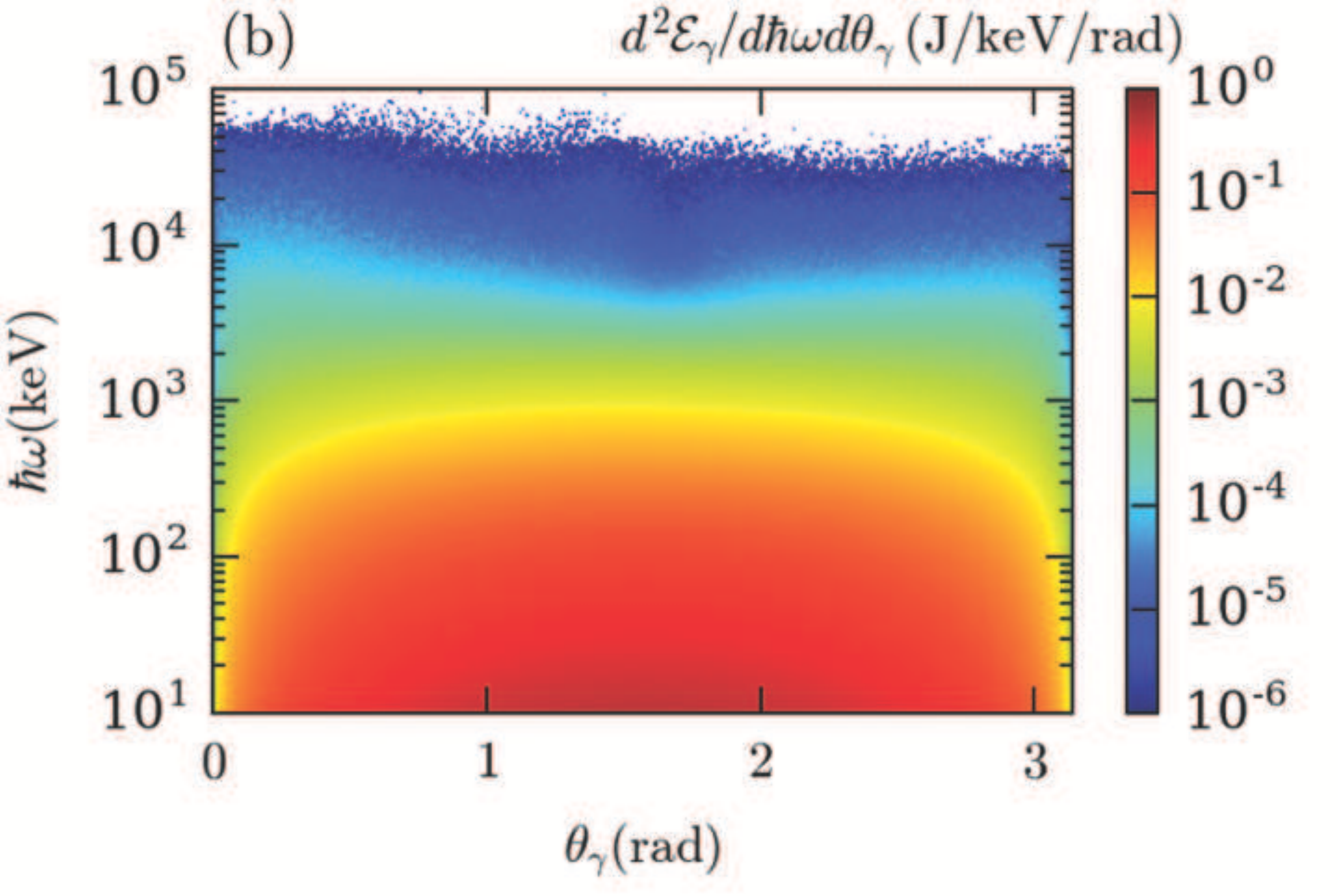}
\caption{Time-integrated (a) synchrotron and (b) Bremsstrahlung energy-angle spectra from the $5\,\rm \mu m$ thick Cu target.
The result is expressed in $\rm J/keV/rad$ instead of $\rm J/keV/rad/\mu m$ as we assumed the third dimension $z$ to be $5 \, \rm \mu m$.
}
\label{fig:dPdthedE_sync_brem_l5mum}.
\end{figure}

The time-resolved synchrotron angular power spectrum is displayed in Fig.~\ref{fig:dPdth_l5mum_2D}(a). The synchrotron emission is observed to peak at $t\simeq 7\,\rm fs$, i.e., just after the on-target laser maximum. The corresponding spatial distributions of the laser field and electron density are shown in Fig.~\ref{fig:bz_ne_pr_l5mum_2D_t07fs_it2800}. Because of the sustained compression of the irradiated boundary, the electron density profile is locally much steeper than at the same time in the fast-expanding $l=32\,\rm nm$ foil (see Fig.~\ref{fig:bz_ne_pr_l32nm_2D_t12fs_it30000}), which, in turn, leads to significantly less energetic electrons (with longitudinal momenta up to $p_x \simeq 150\,m_ec$ vs. $p_x > 400\,m_e c$ at $l=32\,\rm nm$). Such interaction conditions are close to those characterizing the early stage of Fig.~\ref{fig:oc_plasmas_interaction}. Accordingly, synchrotron emission arises in front of the laser-compressed boundary (see red isocontour at $P_\gamma = 10^{23}\,\rm W\,cm^{-3}$ in Fig.~\ref{fig:bz_ne_pr_l32nm_2D_t12fs_it30000}), where $10 \lesssim n_e \lesssim 100\,n_c$, and is mainly forward directed ($\theta_\gamma \lesssim 1.5$). Some backward emission also occurs by $t \simeq 20\,\rm fs$, i.e., after a two-way transit time in the foil of the energetic electrons generated in the laser’s rising edge, but contributes weakly to the total angular spectrum [see Fig.~\ref{fig:dPdthedE_sync_brem_l5mum}(a) and Fig.~\ref{fig:spc_phot_sync_simus2D}(b), discussed below]. 

Let us now examine the Bremsstrahlung spectrum presented in Fig~\ref{fig:dPdth_l5mum_2D}(b). Similarly to the $l=32\,\rm nm$ target, but to a greater extent given the two orders of magnitude larger thickness, a significant ($\sim 10^{10}\,\rm W/rad/\mu m$) isotropic background is radiated early on by the thermal electrons.
The total Bremsstrahlung power increases by a factor of $\sim 10$ during the main part of the pulse ($-50 \lesssim t \lesssim 50\,\rm fs$) and essentially saturates afterwards. In such a thick target, the maximum simulation time ($t=+250\,\rm fs$ after the on-target laser maximum) allowed by our computational resources is clearly too short for a quantitative evaluation of the total Bremsstrahlung yield. Except for this shortcoming, the Bremsstrahlung spectrum at $l=5\,\rm \mu m$ evolves qualitatively as observed at $l=32\,\rm nm$. Just after the laser peak, it is mainly contained in forward and backward lobes and, as time passes, it gets increasingly isotropic due to the growing average isotropy of the electron distribution. 

\begin{figure}[tb]
\centering
\includegraphics[width=0.45\textwidth]{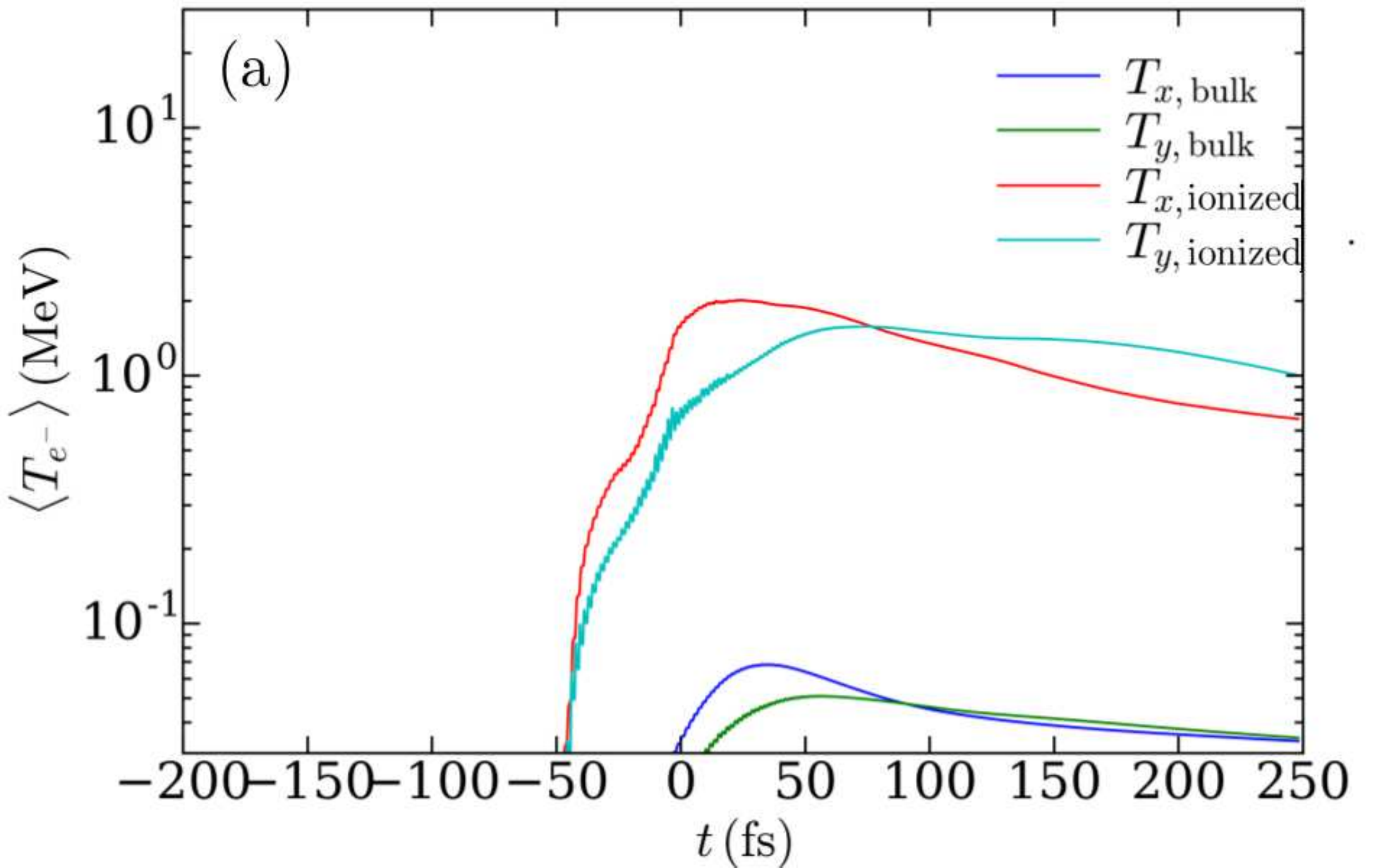}
\includegraphics[width=0.4\textwidth]{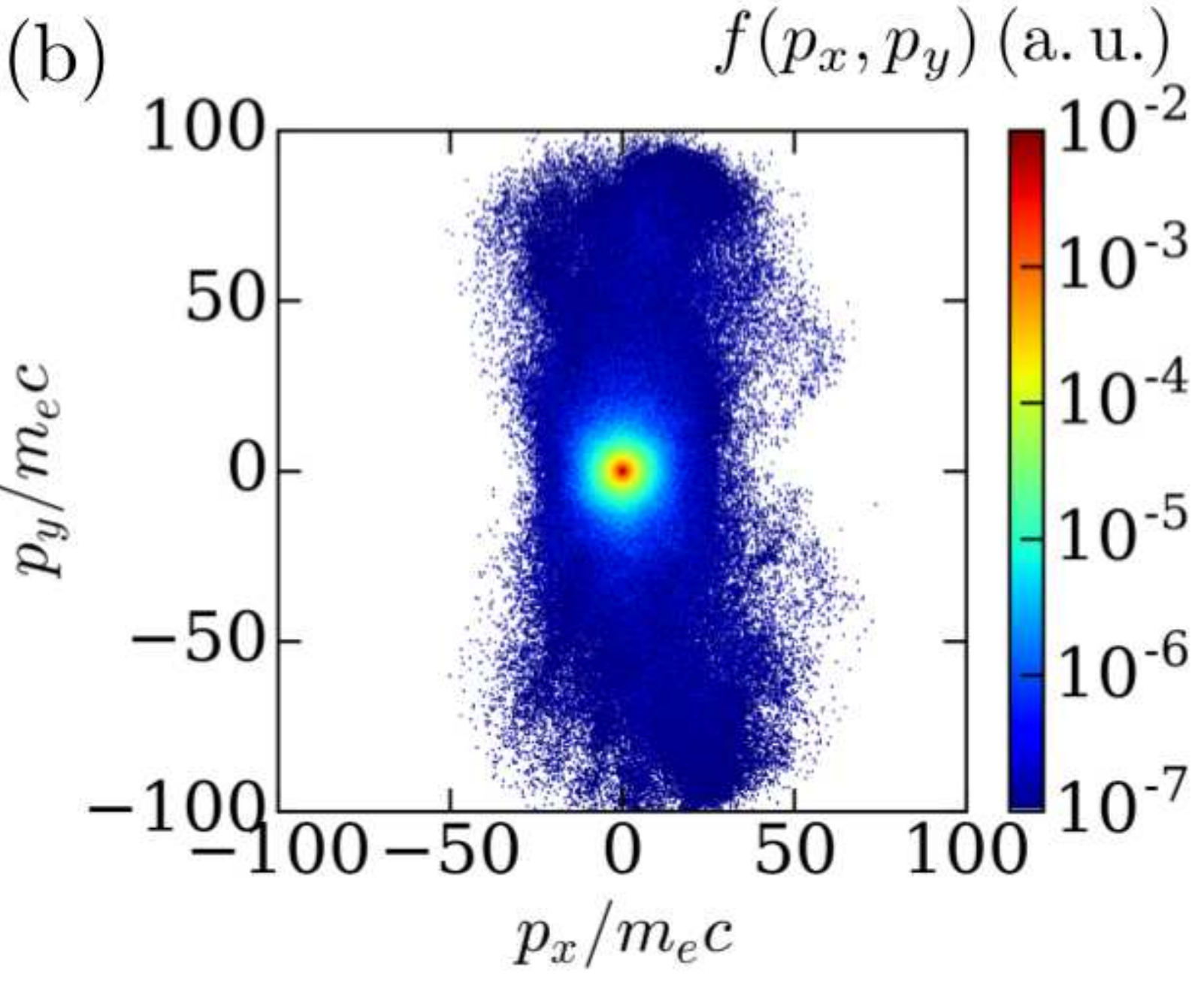}
\caption{(a) Time evolution of the longitudinal ($T_x$) and transverse ($T_y$) electron temperatures in the $5\,\rm \mu m$-thick Cu target. We distinguish between the (`bulk') electrons initially contained in the preionized Cu$^{25+}$ layer and the (`ionized') electrons originating from the surface hydrogen layers and subsequent ionization of the Cu ions. (b) Space-integrated $p_x-p_y$ electron distribution at $t=+250\,\rm fs$.}
\label{fig:tempex_tempey_pxpy_250fs_l5mum}.
\end{figure}

To clarify the electron relaxation dynamics, we have followed the time evolution of the longitudinal and transverse temperatures of two groups of electrons (the sum of which make up the whole electron population): those (`bulk') initially contained in the preionized Cu$^{25+}$ layer and those (`ionized') issued from the surface hydrogen layers and subsequent ionization of the Cu ions. The latter group notably includes surface electrons directly laser-accelerated to high energies, and so reaches much higher temperatures than the former group. The longitudinal ($T_x$) and transverse ($T_y$) temperatures of each group are defined as the momentum fluxes $T_{x,y} = \int d^3p\, f_e(\mathbf{p}) p_{x,y}^2/m_e \gamma$ ($f_e$ is the space-averaged electron momentum distribution). Figure~\ref{fig:tempex_tempey_pxpy_250fs_l5mum}(a) indicates that, as expected, the longitudinal temperature initially grows the fastest for both electron groups. Specifically,  $T_{x,\rm ionized}$ peaks (at $\sim 2\,\rm MeV$) at the laser maximum, after which it steadily decreases down to $\sim  0.7\,\rm MeV$ at $t=250\,\rm fs$. Meanwhile, $T_{y,\rm ionized}$, which is about twice lower at the laser maximum, goes on rising up to $t \simeq 70\,\rm fs$ at which time it overtakes $T_{x,\rm ionized}$ before stagnating/slowly decreasing later on, so that $T_y/T_x \sim 1.4$ at the final time. This anisotropic relaxation is attributed to preferentially longitudinal momentum losses to the expanding ions, and is more pronounced for the higher-energy electron fraction, as evidenced by the $p_x-p_y$ electron momentum distribution at $t = 250\,\rm fs$ [Fig.~\ref{fig:tempex_tempey_pxpy_250fs_l5mum}(b)]. Meanwhile, the ‘bulk’ electrons reach their maximal longitudinal  ($T_{x, \rm bulk} \simeq 50\,\rm keV$) and transverse ($T_{y,\rm bulk} \simeq 30\,\rm keV$) values around $t \simeq 25\,\rm fs$ and $t\simeq \rm 70\,\rm fs$, respectively. Due to collisional scattering off Cu ions, isotropization is reached at $t \simeq 90\,\rm fs$, and is maintained throughout the subsequent cooling of the bulk electrons.

As a result, the time-integrated Bremsstrahlung energy-angle spectrum plotted in Fig.~\ref{fig:dPdthedE_sync_brem_l5mum}(b) shows a nearly isotropic shape up to photon energies $\sim 1\,\rm MeV$. By contrast, the higher-energy photons, which are emitted by highly relativistic electrons, appear to be more collimated in the longitudinal (forward and backward) directions. Figure~\ref{fig:dPdth_l5mum_2D}(b) indicates that this emission mainly takes place within $\sim 50\,\rm fs$ after the laser maximum. Yet the $>1\,\rm MeV$ energy photons carry only a very weak fraction ($\lesssim 1\,\%$) of the total Bremsstrahlung energy [Fig. \ref{fig:dPdthedE_sync_brem_l5mum}(b)]. To conclude this part, we note that the late-time transverse anisotropy of the ultrarelativistic electrons [Fig.~\ref{fig:tempex_tempey_pxpy_250fs_l5mum}(b)] does not lead to a measurable signal because of their much reduced density fraction.

\subsection{Target thickness dependence of the radiation spectra}

\begin{figure}[tb]
\centering
\includegraphics[width=0.4\textwidth]{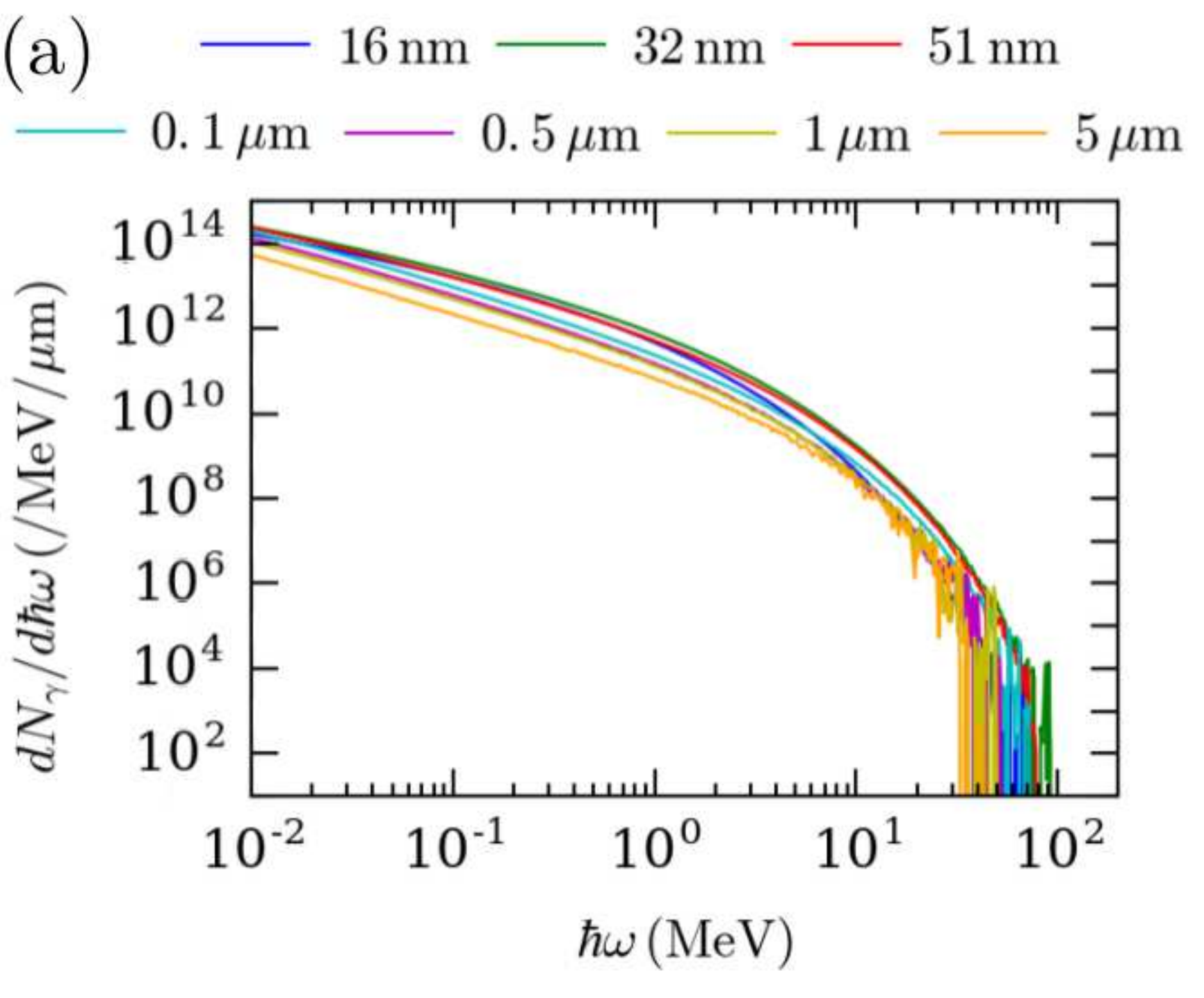}
\includegraphics[width=0.35\textwidth]{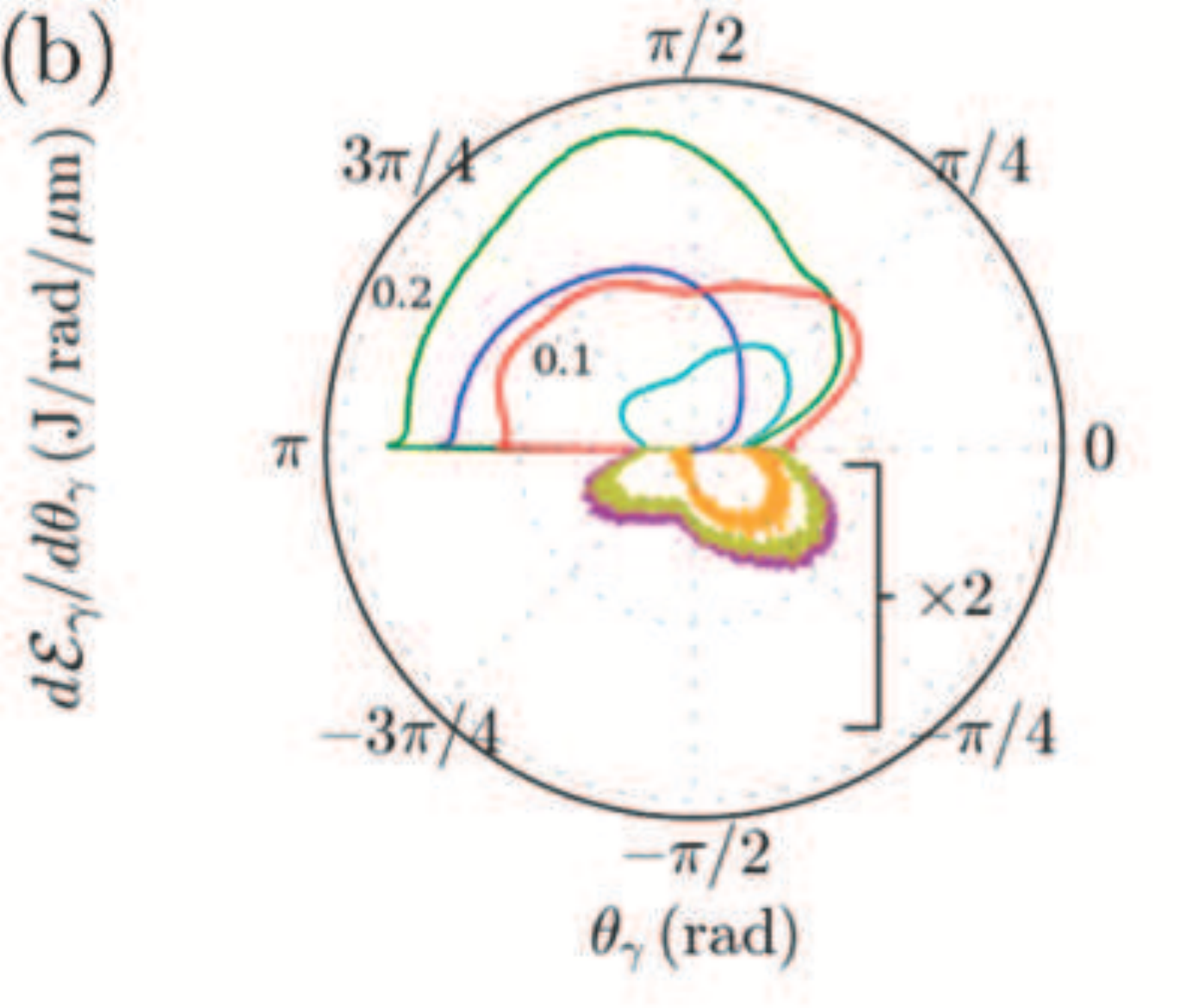}
\caption{Variations in the synchrotron radiation with the Cu target thickness: (a) Energy-resolved and (b) angle-resolved radiated energy spectra for $> 10\,\rm keV$ photon energies. In (b), the energy densities at thicknesses $\le 0.1\,\rm \mu m$ (lower half-plane) have been multiplied by a factor of 2 for visibility.
}
\label{fig:spc_phot_sync_simus2D}
\end{figure}

\begin{figure*}[tb]
\centering
\includegraphics[width=\textwidth]{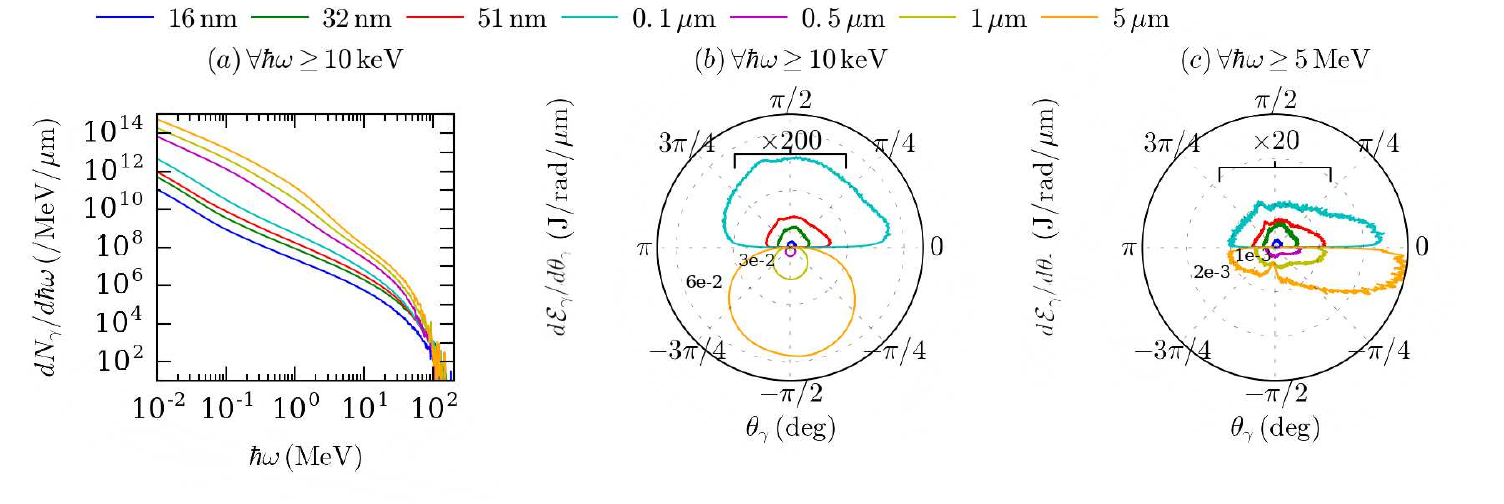}
\caption{Variations in the Bremsstrahlung radiation with the Cu target thickness. (a) Energy-resolved and (b,c) angle-resolved radiated energy spectra for (b) $>10\,\rm keV$ and (c) $>5\,\rm MeV$ photon energies. In (b) and (c), the energy densities at thicknesses $\le 0.1\,\rm \mu m$ (upper half-planes) have been multiplied by a factor of 200 and 20, respectively.
}
\label{fig:spc_phot_brem_simus2D}
\end{figure*}

The properties of the synchrotron and Bremsstrahlung emissions from copper foils of varying thickness are summarized in Figs.~\ref{fig:spc_phot_sync_simus2D}(a,b) 
and Figs.~\ref{fig:spc_phot_brem_simus2D}(a-c). 

The broad energy spectra of synchrotron radiation [Fig.~\ref{fig:spc_phot_sync_simus2D}(a)] exhibit similar monotonically decreasing shapes regardless of the target thickness. They confirm that the maximum yield is achieved at $l=32\,\rm nm$.
The thicknesses $l=32\,\rm nm$ and $51\,\rm nm$ produce the highest photon cutoff energies ($\sim 100\,\rm MeV$), about twice larger than those obtained in micrometric ($l=1-5\,\rm \mu m$) foils.
The synchrotron angular spectra [Fig.~\ref{fig:spc_phot_sync_simus2D}(b)] evidence a clear transition between two distinct angular patterns when the target is made thicker: (i) A dominantly backward/transverse emission at $l=16-51\,\rm nm$, with an oblique forward lobe emerging at larger $l$; (ii) A mainly oblique forward emission at $l=0.5-5\,\rm \mu m$, with a weaker backward lobe, due to refluxing electrons and diminishing at larger $l$.

The Bremsstrahlung energy spectra  [Fig.~\ref{fig:spc_phot_brem_simus2D}(a)] corroborate the growing trend of the Bremsstrahlung yield with the foil thickness as revealed by Fig.~\ref{fig:eff}(a). As also expected from Fig.~\ref{fig:eff}(a), they show stronger variations with $l$ than the synchrotron spectra, across the full range of photon energies. Moreover, they share roughly the same photon cutoff energy ($\sim 100\,\rm MeV$), similar to that of synchrotron emission from nanometric foils.

The Bremsstrahlung angular spectra are displayed in Figs.~\ref{fig:spc_phot_brem_simus2D}(b,c) for two photon groups. The Bremsstrahlung photons with $\hbar \omega \ge 10\,\rm keV$ energies  [Fig.~\ref{fig:spc_phot_brem_simus2D}(b)] are radiated at all angles, but their emission tends to be maximized in the forward direction for $l\le 0.1\,\mu m$ (but the Bremsstrahlung yield is then very weak) while it is essentially isotropic in $l\ge 0.5\,\rm \mu m$ targets (note that an isotropic power spectrum scales with the polar angle as $\sin \theta_\gamma$, as observed at $l \ge 0.5\,\rm \mu m$) due to the dominant contribution of the isotropized moderate-energy ($\lesssim 1\,\rm MeV$) electrons.

Photons with $\hbar \omega \ge 5\,\rm MeV$, on the other hand, are increasingly collimated along the laser axis (and especially in the forward direction) at larger thickness  [Fig.~\ref{fig:spc_phot_brem_simus2D}(c)]. The reason for this trend is that the ultrarelativistic electrons emitting those photons are generated preferentially along the laser axis (coinciding with the target normal), and can recirculate a few times across the solid target (hence the well-defined forward and backward lobes at $l=5\,\rm \mu m$) before losing longitudinal momentum through ion expansion (slowed down at large $l$) or collisions.

\section{Conclusions}
\label{sec:conclusions}

Using advanced particle-in-cell simulations, we have numerically studied the processes of high-energy radiation in plasmas of various kinds, irradiated by laser waves of $10^{22}\, \rm W\,cm^{-2}$ intensity. Following several previously published works, we have first reexamined the energy and angular properties of synchrotron radiation in simplified interaction scenarios, involving laser plane waves of infinite or finite duration and fully ionized, uniform-density plasma slabs of semi-infinite or finite thickness. Our simulations confirm the existence of distinct synchrotron emission regimes depending on the density, and therefore the transparency or opacity, of the driven plasma.

At relativistically undercritical density ($n_e = 17n_c$), the photon emission is mainly caused by energetic electrons counterstreaming against the laser wave. Those electrons are injected at high energies toward the laser source across the laser front, in a time-modulated way due to relativistic Doppler effects. As a result, backward-directed radiation bursts are produced throughout the whole laser-filled volume. While forward emission is also significant in semi-infinite plasmas due to nonnegligible reflected light (interacting with forward-moving electrons), it is found to essentially vanish in rapidly expanding, $1\,\rm \mu m$ thick targets. 

At overcritical density ($n_e = 100n_c$), the photon emission initially occurs in the narrow vacuum region where the electrons are energized, and, at later times, in a more extended region encompassing the skin layer and a fraction of the expanding preplasma. In semi-infinite opaque targets, the radiation is dominated by electrons being rotated back to the target, and thus exhibits a broad maximum at forward angles (around $\theta_\gamma \simeq 1$). In $1\,\rm \mu m$ thick targets, the radiation is enhanced with two forward and backward lobes owing to recirculating electrons.

Secondly, we have investigated the competition of the synchrotron and Bremsstrahlung emissions driven by a $10^{22}\,\rm W/cm^2$ intensity, 50~fs laser pulse focused onto solid copper foils, with thicknesses ranging from a few tens of nm to a few $\mu\rm m$. We have looked in great detail into the dynamics and spectral properties of both radiation processes, and correlated them with the ultrafast evolution of the target.
We have found that the synchrotron efficiency is maximized (reaching a $\sim 1\,\%$ conversion efficiency into $>10\,\rm keV$ photons) in $\sim 30-50\,\rm nm$ thick foils which, owing to relativistic and expansion effects, transition from being opaque to transparent during the laser pulse. In this interaction regime, the synchrotron emission takes place throughout the expanding bulk plasma, and is dominated by ultrarelativistic electrons counterpropagating against the incoming wave. The rapid drop in plasma densities then leads to very weak Bremsstrahlung radiation. As the target is made thicker and opaque to the laser pulse, both hot-electron generation and synchrotron emission get localized around the target front side. The synchrotron spectrum is then mainly forward directed, yet may also feature a backward lobe due to electron recirculation during the laser irradiation. As the target expands more slowly with larger thickness, the energized electrons experience higher average densities, which enables efficient Bremsstrahlung over longer time scales. Bremsstrahlung exhibits stronger variations with the target thickness than synchrotron, and turns out to be the dominant radiation source in Cu targets of $l\gtrsim 1,\rm \mu m$ thickness, with a conversion efficiency reaching the percent level. While most of the Bremsstrahung energy into $\ge 10\,\rm keV$ photons is then radiated isotropically due to the prevailing contribution of relatively low-energy isotropized electrons, its high-energy ($\ge 5\,\rm MeV$) fraction is emitted within increasingly collimated forward and backward lobes.

Our results will be of interest for preparing and interpreting experiments on high-energy radiation from laser-solid interactions at the forthcoming multi-PW facilities.

\begin{acknowledgments}
One of the authors (EH) was supported by the French National Research Agency project TULIMA (ANR-17-CE30-0033-01) and the US Air Force project AFOSR No. FA9550-17-1-0382. The authors acknowledge support by GENCI, France for awarding us access to HPC resources at TGCC/CCRT (Grant No. A0010506129).
\end{acknowledgments}

%

\end{document}